\documentstyle[./aasms4,psfig]{article}  
\begin{document}

\title{ROSAT/ASCA OBSERVATIONS OF THE MIXED-MORPHOLOGY SUPERNOVA REMNANT W28}  
\author{Jeonghee Rho}  
\affil{  
SIRTF Science Center, California Institute of Technology, MS 220-6,  
Pasadena, CA 91125,   
e-mail:rho@ipac.caltech.edu}  
\centerline{AND}   
\author{Kazimierz J. Borkowski}  
\affil{Department of Physics, North Carolina State University, Raleigh,  
NC 27695}   

\received{November 2001}
\accepted{March 2002}
\centerline{To appear in the Astrophysical Journal}

%
\begin{abstract}  
  
We present three sets of ROSAT PSPC and four sets of ASCA observations  
of the supernova remnant (SNR) W28.  The overall shape of X-ray emission in  
W28  is elliptical, dominated by a centrally-concentrated interior   
emission, sharply peaked at the center. There are also partial northeastern    
and southwestern shells, and both central and shell X-ray emission   
is highly patchy.   
  
The ASCA spectra reveal emission lines of  Ne, Mg, Si, and  Fe K$\alpha$  
and continuum extending at least up to 7 keV, showing that X-ray emission in   
W28 is mostly of thermal origin with a hot thermal component.   
We found that spectral variations are present in W28.  
The southwestern shell can be fit  
well by a plane-shock model with a temperature of 1.5 keV and an ionization  
timescale of $1.5 \times 10^{11}$ cm$^{-3}$ s.   
The long ionization timescale combined with a low estimated electron density   
of $\sim 0.2$ cm$^{-3}$ implies SNR age of several $\times 10^4$ yr.  
The low density in the southwest is consistent with the shock breakout   
away from molecular clouds in the north and northeast.  
The northeastern shell, with a lower temperature of $0.56$ keV and a longer   
ionization timescale of $1.7 \times 10^{13}$ cm$^{-3}$ s,   
spatially coincides with the radiative shell delineated by radio and optical   
filaments. But a relatively high temperature   
and a low density of X-ray emitting gas in the northeastern shell indicate  
that we are not observing gas cooling from high temperatures.  
Unlike for the southwestern and northeastern shells, the central  
emission cannot be fit well by a single temperature model, but two  
components with temperatures of  0.6 keV and 1.8 keV are required. The  
long ionization timescales imply that the  gas is close to the  
ionization  equilibrium.   
The low temperature component is  
similar to those seen in other Mixed-morphology SNRs.  
The X-ray luminosity of W28 is $\sim 6 \times 10^{34}$ ergs s$^{-1}$,  
and the estimated X-ray mass is only $\sim 20 - 25 M_\odot$.  
A comparison of W28 with other typical Mixed-morphology SNRs  
reveals significant differences in its X-ray properties; W28 has  
a higher temperature and noticeable spectral variations.

W28 belongs to a class of SNRs considered  
by Chevalier (1999), with a radiative shell interacting with clumpy  
molecular  clouds. X-ray emission at its center is   
a ``fossil'' radiation  
from gas which was shocked early in the evolution of the remnant, and its  
centrally-peaked morphology could have been caused   
by processes such as evaporation, electron thermal  
conduction, and mixing induced by various hydrodynamical instabilities.  
But W28 poses a challenge for existing models of X-ray emission,  
because the evaporation model   
of White \& Long (1991) is in conflict with observations, while the presence of  
temperature variations seems inconsistent with SNR models with efficient   
thermal conduction.   
  
\end{abstract}  

\section{Introduction}  
  
Supernova remnants (SNRs) can be classified into three broad categories  
based on their X-ray and radio morphologies. One class is those in  
which both the X-ray and radio morphologies are shell-like; most of  
shell-like SNRs have thermal   
X-rays, from a plasma heated  by a shock wave  
(such as in Cas A). Some shell-like SNRs such as SN 1006   
emit nonthermal synchrotron X-rays due to radiation from shock-accelerated   
electrons. The second  
class consists of the Crab-like remnants, also called plerions.  The  
X-ray and radio morphologies of these are center-filled. Their energy  
source is an active pulsar (Seward 1985),  and the X-ray spectrum is a  
featureless power law due to synchrotron processes. The third  
morphological category is a composite, 
simultaneously central emission and a shell (such as Vela and CTB 80).  
Recently we have classified a group of SNRs, which have centrally  
peaked X-ray and shell-like radio morphology, as a new, distinct  
morphological class which we term  ``mixed-morphology'' (MM) supernova  
remnants (Rho \& Petre 1998) or ``thermal composite'' SNRs (Jones et al. 1998).   
Members of this class  
include 3C391, W28, and W44. Despite their X-ray morphological similarity to  
the plerions all X-ray spectral evidence now clearly show that the  
X-ray emission is primarily thermal (Rho et al. 1994; Rho 1995;  
Rho \& Petre 1998). ASCA observations revealed  
the presence of strong thermal line emission in many of these  
remnants: W44 (Harrus et al. 1997; Rho et al. 1994),  
Kes 27 (Rho et al. 1998), 
3C391
(Chen \& Slane 2001).  The population of these  
mixed morphology SNRs is at least 10\% of all known SNRs, and more than  
25\% among X-ray detected SNRs, suggesting that the mechanism for  
creating a remnant with the centrally-peaked thermal emission is not rare   
(Rho \& Petre 1998).  
  
The most important physical property of these remnants distinctly  
different from others is that their temperature profiles are uniform,  
rather than radially dependent. A radial temperature gradient is  
generally expected for   
a remnant of an SN explosion in a uniform  
ambient medium, such as in the classical Sedov solution.  
The uniform profile was first detected through  
spatially-resolved PSPC spectroscopy of W44 (Rho et al.  
1994), and IC 443 (Asaoka \& Aschenbach 1994; Rho  
et al. 1995).  
The spectral hardness maps of MSH 11-61A, CTB 1, W63, HB 3 (Rho  
1995) and 3C400.2 (Saken et al. 1995) are smooth, also indicating  
uniform temperature profiles.  
  
There are two competing scenarios which have been proposed to explain  
the centrally-filled morphology of mixed-morphology (MM) remnants:  
fossil radiation from the interior of a remnant in the  
shell-formation (radiative phase) stage (Cox et al. 1999; Chevalier 1999), and evaporation of interior   
clouds (White \& Long 1991). The first hypothesis is {\it radiative model}   
in which the shell of an expanding  
SNR cools below $\sim 10^6$ K, and its X-ray emission becomes undetectable  
because of low temperatures and the ISM absorption. But X-rays from the  
hotter interior are still detectable as  ``fossil'' thermal radiation  
(Seward 1985).   
Although dense environments of MM SNRs could hasten the onset of the   
shell-formation   
evolutionary stage, their uniform temperature profiles still stand in  
contrast to the radial temperature gradient expected from ``fossil  
radiation.'' Recent work by Cox et al. (1999)  
and  Shelton et al. (1999), however, suggests  
that thermal conduction significantly reduces the radial thermal  
gradient and thus it could explain the centrally-condensed X-ray  
morphology and the uniform temperature profiles observed  
in the MM SNRs.  
  
The other hypothesis is the {\it evaporation model}, in which   
the enhanced interior X-ray emission could arise from gas evaporated  
from clouds.  A model based on cloud evaporation (McKee 1982; White \&  
Long 1991) reproduces some of properties of a number of MM remnants  
observed by Rho \& Petre (1998).  
This model assumes SNR shock propagation through a cloudy ISM.  Clouds  
which are too small to affect the blast wave propagation, but of  
sufficiently high density to survive passage through a strong shock,  
subsequently provide a reservoir of material inside the remnant  
cavity.   
  
A correlation  between molecular interacting SNRs  
and X-ray center-filled SNRs   
was suggested by Rho \& Petre (1998).  
Recently Chevalier (1999) presented a framework for the SNR evolution  
in molecular clouds, where the radiative shell is interacting with   
clumpy molecular clouds. In his framework X-ray emission is a ``fossil''   
radiation from the SNR interior, although he admits that the observed   
X-ray morphologies of MM SNRs might pose the greatest challenge   
for his models. Just like in the conduction models of Cox et al. (1999) and   
Shelton et al. (1999),  
thermal conduction might be crucial for explaining uniform temperature  
profiles in interiors of these remnants.  
  
The supernova remnant W28 (G6.4-0.1) is an archetype  
MM SNR, showing center-filled X-rays and  
a shell-like radio morphology (Long et al. 1991).   
There are double radio shells  in the northern part of the SNR, and the  
northeastern and eastern shells are apparent;  the magnetic field lines  
are tangential over the northern shell (Frail, Kassim, \& Weiler  
1994; Shaver \& Goss 1970;  Kundu \& Velusamy 1972).  The radio  
spectral index is $\alpha \sim 0.4$ (Kassim 1992), where $S \propto$  
$\nu^{-\alpha}$. The {\it Einstein} IPC observation revealed that the X-ray   
emission is thermal with a temperature of $2.3 \times 10^7$ K (2 keV)  
and an intervening column density $N_H \sim 2 \times 10^{21}$  
cm$^{-2}$ (Long et al. 1991). Analysis of EXOSAT data also led to a high   
temperature of $2.5 \times 10^7$ K and  
a strong Fe K$\alpha$ line at 6.7 keV was detected (Jones 1991).   
Torii et al.\ (1996) from their preliminary analysis of ASCA data  
reported that a combination of thermal and nonthermal models  
provided a good fit to X-ray spectra of W28.  
  
Diffuse optical nebulosity was detected throughout most of the  
interior of W28 (van den Bergh 1978). W28 lies in a complicated region  
of the inner Galaxy which is  confused by several large \ion{H}{2} regions.  
Wootten (1981) suggested an interaction with a  
molecular cloud in W28, based on observations of broad CO (1-0) lines with  
FWHM of 11 km s$^{-1}$ and the presence of warm dense clouds. A recent  
convincing detection of even broader lines (as large as 70 km s$^{-1}$) was  
reported by Arikawa et al. (1999) and Frail \&\ Mitchell (1998) using   
CO (3-2) line. Arikawa et al. mapped shocked gas at the northern  
boundary, finding about 2000 $M_{\odot}$ of shocked molecular gas.  
Twenty-six shock-excited OH masers which are signposts of molecular   
interactions were also   
detected along the northern boundary and along the eastern   
shell of the remnant (Frail, Goss, \& Slysh 1994; Claussen et al. 1999).   
In addition, the  
infrared H$_2$ S(3) and S(9) lines were seen with ISO (Reach \& Rho 2000),  
together with strong atomic fine-structure infrared lines (such  
as C, N, O, Si, P and Fe), with the [\,O I\,] line ratios implying   
a high ($> 10^3$ cm$^{-3}$) medium density. All these observations  
provide an unambiguous evidence for interaction with molecular clouds in W28.  
The presence of molecular clouds visibly affects the
overall shape of the remnant, which is far from uniform and circular, typical
for SNRs; it is instead dented in the north and east where molecular clouds are
located.

W28 has long been considered as a highly evolved remnant (e. g., see   
Lozinskaya 1992).   
In the north, gas behind the blast wave propagating through the interclump   
medium of molecular clouds collapsed into a radiative shell, which can  
be observed as filamentary emission at optical and radio wavelengths.    
Optical spectra typical of radiative shocks, the tangential orientation of the   
magnetic field (Milne \& Dickel 1975), and  
an excellent correspondence between optical and radio emission (Dubner   
et al. 2000) provide irrefutable evidence that the northern half of W28 is   
in the late radiative stage of the SNR evolution. The weakness of the observed  
X-ray emission at the position of the northern radio shells adds further  
support for this conclusion. The blast wave most likely  
broke out of the molecular cloud in the south (Dubner et al. 2000).  
  
Two candidates for a pulsar associated with W28 have been suggested  
(Andrews et al. 1983; Kaspi et al. 1993).   
A glitching pulsar PSR  
B1758-23 was detected at the northern boundary of W28 (Kaspi et  
al. 1993; Frail, Kassim, \& Weiler 1994).  
This remnant might also be the  
gamma-ray source 2CG 006-00 (Ormes 1988; Esposito et al., 1996), and it might harbor a recently   
discovered soft gamma ray repeater, SGR 1801-23 (Cline et al. 2000). 

The distance to W28 is uncertain, between 1.6--3 kpc (Clark \& Caswell  
1976; Milne 1979; Frail, Kassim, \& Weiler 1994).   The $\Sigma$-D relation  
and optical measurements suggest 1.8 kpc (Goudis 1976) and 2  
kpc (Long et al. 1991), respectively. Based on the assumed association of   
PSR 1758-23 with W28 (Kaspi et al. 1993), a distance of 3 kpc was suggested  
using an H I absorption measurement toward this pulsar (Frail, Kulkarni, \&  
Vasisht 1993).   However, this association was not confirmed   
by VLBA measurements of interstellar scattering toward the pulsar   
(Claussen et al. 1999).  Recent CO observations of the shock interaction  
with the molecular cloud imply a distance of 1.8 kpc, adopting the rotation   
curve of   
Clemens (1985) for a galactocentric distance of the Sun of 8.5 kpc.  
Because the evidence for the interaction with molecular clouds is very   
convincing, we adopt 1.8 kpc as the distance to W28.

In this paper, we present ROSAT and ASCA images and  
spectral mapping of W28 using three sets of ROSAT and  
four sets of ASCA data, and we  
compare them to  
other MM SNRs.   
We interpret the results in the framework of an SNR evolution in   
molecular clouds.   
We also examine other nearby X-ray sources in the field of view and   
search for X-ray emission from the pulsar PSR B1758-23.

\section{Observations}  
  
W28 was observed with the ROSAT Position Sensitive  
Proportional Counter (PSPC). The count rate is 5.78$\pm0.11$ cts s$^{-1}$,  
almost twice as much as the IPC count rate of 3.2 cts s$^{-1}$ (Seward 1990).  
A total of three pointed observations were performed, with  
the observation parameters listed in Table \ref{w28obser}.   
The total  
exposure time is 23,800 s. The 45$^{\prime}$ remnant size extends  
beyond the central circular field of view   of the PSPC, where the PSPC response  
is most uniform and sensitive. Therefore, an exposure correction was  
necessary in order to obtain the mosaiced image. We used  
the analysis techniques for extended objects developed by Snowden et  
al. (1994).   
The scattered solar X-rays,  the time-dependent flux of short- and  
long-term enhancements, and background contamination were evaluated as  
three parameters of the total number of contaminating counts in each  
band, the gradient of the counts across the image, and the rotation  
angle of that gradient, and they were subtracted. Residual  
afterpulse events were also removed.  The vignetting  and  
detector-artifact corrected exposure maps, and modeled particle  
background maps (including the soft background using a background model)  
were created and applied to the data.  After correction of three images   
for background and  
exposure time, the images were merged together,  and smoothed using   
adaptive-filtering algorithm.

We used archival ASCA data in order to investigate  
spectral properties of W28 at moderate  ($E/\Delta E = 60$ at 6 keV) spectral  
resolution. ASCA (Tanaka, Inoue, \& Holt 1994) had two  
detector pairs: Gas Imaging Spectrometers (GIS2 and GIS3) and  
Solid-state Imaging Spectrometers (SIS0 and SIS1).  The SIS covers an  
energy band of 0.5--10 keV and the GIS 0.6--10 keV.  The on-axis  
angular resolution of the GIS and SIS is about 1--2 arcminutes. Four  
sets of ASCA data pointed toward the northeast, center (two sets of data) and   
southwestern portion of the remnant were used, and the observational  
parameters are summarized in Table~\ref{w28obser}.  
We filtered the ASCA data using several criteria such as Cut off Rigidity  
(COR) and earth elevation (based on revision 2 processing).     
  
We generated two ASCA mosaiced images in 0.6--10.0 keV and  
hard (3--10 keV) energy bands.    
We used 4 sets of  GIS2 and GIS3 images  
with exposure correction and background subtraction (the GIS fields of  
view are marked in Figure \ref{w28ascafov}).   
First, the instrument maps were generated using FTOOLS task ``ASCAEFFMAP",  
and the  
effective exposure maps were generated by combining the instrument maps  
with exposure times (``ASCAEXPO"). To subtract the background, we   
used long-exposure blank sky observations from which  
all point-like sources brighter than $2 \times 10^{-13}$ ergs s$^{-1}$  
cm$^{-2}$ in the 2--10 keV range were removed. These files were  
generated using high latitude observations taken between June 1993 and  
December 1995 by the ASCA team (Ikebe et al. 1998). The outer detector edges   
beyond 17$'$ radius were removed.    
Then we mosaiced 4 sets of GIS images using ``FMOSAIC".

\section{Results}  
  
\subsection{Images and Morphology}

The mosaiced ROSAT PSPC image is shown in Figure~\ref{w28pspc}.   
The coverage of the GIS ASCA observations is marked on the  
ROSAT PSPC image in Figure \ref{w28ascafov}. The mosaiced ASCA image is  
shown in Figure \ref{w28ascaall}, with the ROSAT contours superposed.  
The ASCA image is consistent with the ROSAT image, considering that  
its spatial resolution is much  
lower, only about 1\arcmin, and that it also includes photons with energies  
higher than 2 keV which could not be detected by ROSAT.  
The global shape is elliptical; X-ray emission  
is concentrated at the center and  fills  the entire interior of the  
remnant. The PSPC image reveals that the central  
emission is  sharply peaked and highly  
patchy.   
Limb-brightened partial shell  
structures are present in the  northeast and southwest, and the  
northeastern shell is also partially  broken, with patchy structures.   
There are a few places  showing highly structured X-ray  
emission: there are ear-like segments of limb-brightened northern shell  
and a sudden jump (northern flat boundary in Fig.~\ref{w28regionmark}) in surface brightness in the north.  
The X-ray surface  
brightness  between the two ear-like patches (left patch is the northeastern shell  
and right patch is shown in Fig.~\ref{w28regionmark}) stays very flat from the  
east to west along Dec -23$^{\circ}$ 15$'$.  The PSPC image is  
superposed on contours of a 328 MHz radio continuum image (Dubner et al. 2000)  
in Figure \ref{w28xrayradio}. The X-ray surface brightness is  
anticorrelated with radio morphology not only on  large scales, but  
also on small scales: the eastern prominent radio shell is located outside  
the X-ray emission and the northern thin inner radio shell is immediately  
outside the northern flat boundary of X-ray emission. In the north,  
double radio shells are present, which is probably due to projection,  
 and in the southwest both X-ray and radio  
emission are faint but show a partial shell. The only places where the radio  
and X-ray emission match well are the two ear-like structures, reinforcing  
the suggestion that the emission here is from the shell.       The  
northeastern shell matches well with the partial radio shell  as shown in the  
radio superposed image.   Faint emission is found   in the southwest,  
apart from the main body of the remnant.  This part of the SNR is seen  
faintly in X-ray and radio  but not at optical wavelengths.  This suggests that  
it is either a breakout region of the remnant, or that extinction is higher here.  
The X-ray emission may extend beyond the radio boundary in  
the west.

In the PSPC soft band ($E < 0.5$ keV) most of the emission arises  from  
the central part of the remnant and a bright patch in the  northeastern  
shell, as shown in Figure~\ref{w28soft}.   
We created  
a PSPC hardness map between energy band of 0.9--2.2 keV  and  
0.5--0.9 keV (i. e., bands 5--7 and bands 2--4  as defined by Snowden et  
al. 1991).  We chose the energy boundary of 0.9 keV because  
the spectral peak of the PSPC spectrum appears at this energy (see   
Figure~\ref{w28spec}). A spectral hardness ratio (SHR) map, defined here as the ratio of the  
high energy band map to the low energy band map, can reveal  
absorption and/or temperature variations.  A high value of SHR  
indicates a locally high column density or a high temperature.   
The hardness ratio map is superposed on the PSPC surface  
brightness contours in Figure~\ref{w28shr}.    
This hardness map shows that the emission  
is harder in the southwest and south than in the north and  
northeast. This is consistent with our  spectral results   
that the southwestern shell has a higher temperature than  
that of northeastern shell, and possibly elsewhere in the remnant    
(see \S~3.2).  
We generated an ASCA hard (3--10 keV)  
image as shown in Figure \ref{w28ascahard};  
the southwestern part is brighter  
in hard X-rays relative to the total energy ASCA map. The central  
bright peak is also bright at hard X-rays, and the hard  
energy maps give hint of  
hard emission at the patches of both ear structures and the northern  
boundary.  A hard source appears in the lower right corner of the image in Figure \ref{w28ascahard},
which position coincides with two IRAS sources and an A star (HD 313524).

In order to show clearly differences in the radio and X-ray morphologies,  
we compare X-ray and radio surface brightness profiles in Figure~\ref{w28sb}.   
We chose  
the strong X-ray peak ($\alpha$ = 17$^{\rm h}$ 57$^{\rm m}$ 23.2$^{\rm s}$,  
$\delta$ = -23$^{\circ}$23$^{\prime}$50.6$''$, B1950) 
as the center for this analysis.   
We divided the remnant into 5 sectors based on their  
different appearance in radio and X-ray images: NW, covering position angles   
between  
285$^{\circ}$ - 360$^{\circ}$ (measured east of north);  
N, covering P.A. 0$^{\circ}$ to 45$^{\circ}$ and 250$^{\circ}$ to   
360$^{\circ}$;   
SW,   
185$^{\circ}$ to 250$^{\circ}$; SE, 85$^{\circ}$ to 160$^{\circ}$;  
and northeast, 65$^{\circ}$ to 70$^{\circ}$.  
All X-ray profiles peak at the center, while the  
radio profiles show a prominent shell in most directions, but weakest at  
the south and southwest. A noticeable feature of the X-ray surface  
brightness is that it shows a shell structure in the northeast  
which coincides with the radio shell at  
20\arcmin\ distance from the center (see Figure~\ref{w28sb}). 
  
The surface brightness is composed of 3 components: 1) innermost central part   
with centrally peaked brightness ($r_s <  2$\arcmin),  
2) smooth component (2\arcmin\ $< r_s <$ 35\arcmin), and 3)  northeastern and   
southwestern shell   
(20--30\arcmin\ in radius) components.  
In the innermost central region, the spectrum  
is rather hard, and this region can be seen in both hard and soft  
emission in Figure~\ref{w28ascahard}.   
The observed surface  
brightness distribution in the interior of W28 implies a radial density   
profile described by  
$n/n_c = 1 - r/R$, as shown in Fig. \ref{w28sb}a,   
where $n(r)$ is plasma density as a function of  
distance $r$ from the remnant's center, $n_c$ is the central density,   
and $R$ is the remnant's radius. This profile is  
steeper than profiles usually found in other mixed-morphology SNRs  
(Rho \& Petre 1998). The overall  
shape of remnant is far from uniform and circular, typical for supernova  
remnants; it is dented in the north and east because of the presence of   
molecular clouds. Partial shell structures and X-ray features in  
the interior are often adjacent to molecular interaction sites, and appear  
particularly bright when projection effects allow for longer   
line-of-sight distances through the X-ray emitting gas. This most often   
happens in regions least affected by the irregular cloud geometry, such as  
the northeastern shell. The projection effects are not  
favorable in the north, where molecular clouds are located on the near   
(front) side of the remnant (Arikawa et al. 1999). Instead of a shell,  
we observe an abrupt drop in the X-ray surface brightness at the location of  
molecular clouds (the northern flat boundary).   
  
\subsection{Spectral Analysis using ASCA/ROSAT spectra}

We extracted three sets of ASCA and ROSAT PSPC spectra  
for   
the northeastern shell, center and southwestern shell (see  
Figure~\ref{w28regionmark}).   
The GIS and SIS spectra  
were extracted from each of the three pointings (we used the long exposure of  
sequence 51022000 for central region).  
The central region spectra covered most  
of GIS and SIS fields of view in the central pointing   
(ad51022000; see Figure \ref{w28ascafov}).   
The southwestern shell and the northeastern shell extraction regions follow   
the shapes of the shells as shown in Figure \ref{w28regionmark}:   
for the southwestern  
shell, the extracted region is  
a 12$' \times 8'$ ellipse inclined at PA 162$^{\circ}$,  
and for the northeastern spectra, it is an 8\farcm2 $\times$ 3\farcm2 ellipse at   
PA 150$^{\circ}$.  
The   
central spectra, after background subtraction, have count rates of   
1.004$\pm$0.007 cts s$^{-1}$  for GIS2,  0.734$\pm$0.006 cts s$^{-1}$ for GIS3,   
0.785$\pm$0.007 for SIS0, 0.950$\pm$0.008 cts s$^{-1}$ for SIS1,   
and 0.704$\pm$0.015 cts s$^{-1}$ for ROSAT PSPC.   
The southwestern shell spectra after background subtraction have count rates of  
0.412$\pm$0.006 cts s$^{-1}$  for GIS2,  0.320$\pm$0.006 cts s$^{-1}$ for GIS3,  
0.222$\pm$0.006 for SIS0, 0.174$\pm$0.005 cts s$^{-1}$ for SIS1,   
and 0.765$\pm$0.015 cts s$^{-1}$ for ROSAT PSPC.  
Constraints imposed by the internal calibration source led to a reduction  
in the effective GIS3 extraction area, resulting in lower GIS3 count  
rates relative to GIS2.  
The northeastern shell spectra after background subtraction have count rates of  
0.215$\pm$0.003 for SIS0, 0.276$\pm$0.005 cts s$^{-1}$ for SIS1,   
and 0.357$\pm$0.007 cts s$^{-1}$ for ROSAT PSPC. We did not use GIS spectra  
because the signal-to-noise ratio for this small region was too low  
to be useful.  
  
The central spectra show emission lines of Ne, Mg, Si,  
and Fe K$\alpha$, and continuum at least up to 7 keV (Figure \ref{w28spec}a).  
The presence of emission lines means that the  
emission is  thermal   despite the center-filled morphology,  
implying that X-rays are not powered by the pulsar like in the Crab Nebula.   
A single power law fit is significantly rejected.  
This spectrum is different  
from another archetype mixed-morphology SNR, W44, where no emission  
above 5--6 keV is reported. We examined the  
source-free background region to see if the Fe K$\alpha$ line comes from   
the Galactic Ridge, but there is no evidence of Fe K$\alpha$ in source-free  
background spectra. In addition, the Fe K$\alpha$ line image confirms that the  
emission is from the remnant itself, primarily from the center and  
the southwestern shell.   
  
When we compare the central spectra with the southwestern and northeastern  
shell spectra, we find noticeable differences.   
The southwestern spectrum (at RA 17$^{\rm h}$56$^{\rm m}$,   
Dec -23$^\circ$35$'$, see Fig. \ref{w28regionmark}) is much harder than the   
central spectrum as it can be seen in Figure \ref{w28spec}a, while  
the northeastern shell spectrum (see Figure \ref{w28spec}b) is softer  
than spectra in other regions of the remnant. In order to quantify these  
differences, we simultaneously fit the spectra to five sets of  
ROSAT/PSPC, ASCA/SIS0 and SIS1, and ASCA/GIS2 and GIS3 spectra.    
We used  the collisional ionization equilibrium (CIE)   
thermal model (Mewe-Kaastra plasma model; Kaastra 1992), and nonequilibrium  
ionization (NEI) thermal models as implemented in XSPEC  
(Borkowski, Lyerly, \& Reynolds 2001), which included updated Fe L-shell  
atomic data based on calculations by Liedahl, Osterheld, \& Goldstein (1995).   
Fit parameters are tabulated in Table \ref{w28ascafit}.  
The SW spectrum is fit well by the plane shock model with a high  
temperature of $kT_s = 1.2$ keV. A fit with a simpler single ionization  
timescale NEI model reproduced the spectra   
moderately well ($\Delta \chi^2 \sim 1.16$). The Sedov  
model for SW region yielded $kT_{s} = 1.2$ keV, and an   
ionization timescale of $n_et = 3.6 \times 10^{11}$  
cm$^{-3}$ s. The  
northeastern shell is fit best by NEI  
model with a lower temperature of $kT_e = 0.56$ keV, and  
a longer ionization timescale of $n_et = 1.7 \times 10^{13}$  
cm$^{-3}$ s, than those from  the southwestern shell.  
The line-of-sight absorption toward the northeastern shell is   
$N_H = 0.8 \times 10^{22}$ cm$^{-2}$.  
  
For the central spectra,  
one-temperature CIE or NEI models did not   
yield acceptable fits to  
the central spectra ($\Delta \chi^2 > 2.3$, and $\Delta  
\chi^2 > 1.8$, respectively).  
To reproduce  the moderately strong Fe K$\alpha$ line in  
the central region  requires a second thermal component with a higher  
temperature (see Figure \ref{w28spec}c).  
For the two-temperature model fitting, we used three steps in  
spectral fitting. 
First, we fit the Fe K$\alpha$ line centroid using a Gaussian model,  
arriving at 6.6$^{+0.04}_{-0.10}$ keV, which allowed us to estimate  
ionization parameter. 
As the ionization timescale $\tau$ increases, Fe ions become  
more ionized, with electrons gradually removed from the L shell   
which results in a higher line emissivity and a line  
centroid shift to higher energies.  
For $\tau <  10^{10}$ cm$^{-3}$ s, Fe ions are very  
underionized as compared to ionization equilibrium, and the Fe  
K$\alpha$ line is produced mostly through fluorescence following  
K-shell ionization by electrons. 
The observed  centroid  implies an ionization time scale for the hot  
temperature component of $\sim 5 \times 10^{11}$ cm$^{-3}$ s (e. g., see Fig.  
2 of Borkowski \& Szymkowiak 1997).  As the second step, we  constrained the   
temperature and  
Fe abundances using only hard energy band (4--10 keV) spectra, with  
the bremsstrahlung model for the continuum and a Gaussian line    
for Fe K$\alpha$. The constraint  
is 1.8 keV for the central  
region, and the Fe abundance is  consistent with solar abundances.  
A single ionization timescale NEI model fit in the hard energy band  
alone also gave $\tau = 5 \times 10^{11}$ cm$^{-3}$ s, 1.8 keV temperature,  
and approximately solar Fe abundance.  
The fits are summarized in Table \ref{w28ascafit}. As the third step,  
we added a second, soft component (CEI  
model, single ionization timescale NEI model, or Sedov  
model), while the temperature and  
ionization timescale for the hot component were fixed to the values   
obtained in previous steps.   
A  two-temperature  
NEI model   
reasonably well reproduced the observed  
spectra  ($\Delta\chi^2 \sim 1.1$, see Figure \ref{w28spec}d) with a hot   
temperature of $kT_{hot}  
= 1.8\pm0.5$ keV and ionization timescale of  
$\tau_h = 5 \times 10^{11}$ s~cm$^{-3}$, and a low temperature of  
$kT_l = 0.6$ keV with $\tau_l = 2 \times 10^{12}$ s~cm$^{-3}$.  
(We note that modeling with just 2  
temperature  components is merely a convenient way of quantifying  
temperature differences present in the remnant. Because  
multi-temperature plasmas might be expected in complex remnants such as W28,   
a considerable caution is required in  
interpreting results of the two-component fits.)   
Both NEI and Sedov models provide equally good fits for the soft component,  
with a high ionization parameter of $\sim 2 \times 10^{12}$ cm$^{-3}$ s.   
  
Torii et al. (1996) suggested the presence of nonthermal X-ray emission in   
the southwest. With the poor S/N SW spectra, the presence of nonthermal   
emission is still an open question, which cannot be completely ruled out   
at this time.  
But because we see Fe K$\alpha$ line in GIS2 spectra of the SW region, we   
favor thermal over nonthermal origin for the high energy emission.   
If nonthermal component were present in SW spectra, we would have  
a combination of thermal and nonthermal emission, because thermal line   
emission from elements such as Mg and Si is certainly present at low  
energies.  
It is also possible that the hard X-ray continuum and the Fe K$\alpha$   
line come from different regions.   
We attempted to find differences in spatial distributions between the  
hard continuum and the Fe K$\alpha$ line, but this attempt  
failed because of low count rates at high energies.  
Better data are needed to confirm that the southwestern   
emission is of purely thermal origin, and to study the remnant's  
morphology in the Fe K$\alpha$ line and in the hard continuum.  
  
The derived ISM absorption is higher in the SW than in the northeastern  
shell and the central part of remnant (see Table 2). Considering that  
the remnant is interacting with clouds in north, east, and northeast,  
one might expect an enhanced absorption in these regions of the remnant,  
just the opposite of what is observed. However, if  
there are clouds in front of the remnant in the SW, the line-of-sight  
absorption can still be high  
without any interaction between the remnant and clouds.  
This is the case of IC 443, where $N_H$ is higher across the central part of  
remnant because of the presence of  
molecular clouds in the front of the remnant, although the interaction  
with the clouds is seen in the south (Rho, Petre, \& Hester 1994).  
There is  
no optical emission detected in the SW part of W28 (van den Bergh 1978;  
Long et al. 1991; Winkler, in private communication) which seems to suggest  
that absorption is indeed higher in the SW. But an equally plausible  
explanation  
for the weakness of optical lines is simply the absence of radiative shocks  
in the breakout SW region of the remnant. More observations are needed in  
order to resolve this issue, because the poor S/N spectrum of the  
southwestern shell ultimately limits our ability to understand the origin  
of its relatively hard spectrum.

\subsection{Other sources in the field of view including the pulsar}  
  
A glitching pulsar, PSR B1758-23, was detected at the northern  boundary of  
W28 (Kaspi et al. 1993; Frail, Kassim, \& Weiler 1994), at  
R.A. $18^{\rm h} 01^{\rm m} 19.859^{\rm  
s}\pm0.058^{\rm s}$   
and Dec. $-23^\circ$ 06$^{\prime} 17^{\prime \prime}$$\pm 102''$.  
An association of this pulsar with W28 was  
suggested by Kaspi et al. (1993), with   
the pulsar distance estimated at  
3 kpc (Frail, Kulkarni, \& Vasisht 1993). We   
examined this region in detail and found that the PSPC surface brightness  
is 2.5 times higher than the background brightness. 
The pulsar position is marked on  
the PSPC image in Figure~\ref{w28pulsar} (rp900395 set of data, 9.3 ks in duration).   
Within errors, the centroid coincides with the pulsar.  
The small number of detected counts was not sufficient to perform a  
spectral or a timing analysis.  
The X-ray PSPC count rate is 3.05$(\pm1.21)\times$10$^{-3}$  
counts s$^{-1}$, which implies a luminosity of  
1.2$\times$10$^{32}$ ergs s$^{-1}$, assuming   
a line-of-sight absorption of $8 \times 10^{21}$ cm$^{-2}$  
and a power law spectrum with $\Gamma = 2$ (equal to the spectral index of   
the  Crab Nebula). The luminosity  
is less than 0.5\% of the total X-ray luminosity of W28.  
A deeper observation is  
required to confirm whether this is true detection or not.

There are two X-ray sources to the east of the remnant. The  
northern source (R.A.  
$18^{\rm h} 02^{\rm m} 10^{\rm s}$ and Dec. $-23^\circ$ 33$^{\prime}  
38^{\prime \prime}$) is very soft, without  
a known counterpart. It is  
probably a stellar source, which we denote as RX1802-2333.  The  
second (southern) source,  
located at R.A. $18^{\rm h} 01^{\rm m} 59.7^{\rm s}$ and Dec.  
$-23^\circ$ 41$^{\prime} 20.8^{\prime \prime}$, is the HII region BFS 1,  
and its O-type star, CPD-23 6751, emits X-rays. 
There is another  
X-ray source  in the northeast which can be identified with another H\,II region,  
the Trifid nebula,  
whose observation (R.A. $18^{\rm h} 02^{\rm m} 23.35^{\rm s}$ and Dec.  
$-23^\circ$ 01$^{\prime} 47.0^{\prime \prime}$) is reported elsewhere   
(Rho et al. 2001).

\section{Discussion}  
  
Because W28 is a complex remnant in an advanced evolutionary stage, we first   
discuss its global kinematics and dynamics based on  
observations at wavelength bands other than X-rays. Next, we discuss and   
interpret X-ray observations. Finally we compare W28 with other   
mixed-morphology remnants.  
  
\subsection{Kinematics and Dynamics of W28}  
  
Fabry-Perot observations of the brightest optical emission at the center of  
W28 gave expansion velocity of 40 km s$^{-1}$ (Lozinskaya 1974), with  
individual filaments moving with radial velocities up to 80 km s$^{-1}$ and  
with the average H$\alpha$ width of 40--50 km s$^{-1}$.  
This should also be the blast wave velocity if this optical emission   
projected near the center of W28   
has the same origin as the radiative filaments seen at the northern edge of  
the remnant. Optical emission in the interior has a more diffuse, chaotic  
appearance than the filamentary emission usually associated with radiative  
shocks seen edge on. Such diffuse emission is generally seen in remnants  
at late stages of their evolution (Lozinskaya 1992), and might potentially  
be associated with other physical processes such as cloud evaporation  
(Long et al. 1991) or cooling of X-ray emitting gas in the remnant's  
interior. But at least some of this emission must be produced by   
the radiative blast wave seen in projection against the center of the remnant,  
in view of a generally good correlation between radio, optical, and molecular   
emission (Dubner et al. 2000). Kinematic information about the blast  
wave velocity obtained from Fabry-Perot observations is in a fair agreement   
with estimates of shocks speeds derived from optical spectroscopy:   
Bohigas et al. (1983) estimate shock velocities between 60 and 90 km s$^{-1}$  
from the ratio of [\,O III\,] $\lambda$5007 /H$\alpha$ line strengths,   
while Long et al. (1991) derive velocities larger than 70 km s$^{-1}$  
from ratios of [\,N II\,] and [\,S II\,] line strengths to the H$\alpha$ line  
strength. All available evidence based on optical observations suggests   
that the blast wave velocity $v_s$ is $\le 100$ km s$^{-1}$, with considerable  
spatial scatter with velocities varying in the 60 -- 100 km s$^{-1}$ range.  
  
Only upper limits on the preshock density $n_0$ ahead of the blast   
wave can be   
derived from optical and infrared observations. Long et al. (1991) obtained  
an upper limit for the electron density of 70 cm$^{-3}$ from observations of   
the density sensitive  
[\,S II\,] $\lambda$6717/$\lambda$6731 line ratio. This is consistent with  
the density of $\sim 100$ cm$^{-3}$ derived by Bohigas et al. (1983) from  
this line ratio. An upper limit  
of $\sim 100$ cm$^{-3}$ from the [\,O III] 52$\mu$/88$\mu$ line ratio  
(Reach \&\ Rho 2000) is also consistent with this result. An average preshock   
density must be significantly lower than these upper limits, but by   
an uncertain factor which depends primarily on the magnitude of the magnetic   
field support. In addition, the observed optical emission may originate not   
only in the postshock cooling region. For example, there is some evidence for   
the presence of a faint H II region from variations in the optical line ratios   
across the remnant (Long et al. 1991). Neglecting contribution from  
a suspected H II region and in the absence of the  
magnetic field support, the inferred preshock density would be below  
1 cm$^{-3}$.   
  
Magnetic field is likely to be dynamically important in the  
shocked gas at temperatures where [\,S II\,]  
$\lambda$6717/$\lambda$6731 emission is produced.  
Magnetic pressure becomes  
equal to the gas pressure in the postshock cooling gas at densities  
$n_m = 240 v_s (n_0/10 {\rm cm}^{-3})^{3/2} (B_0/10^{-5}{\rm G})^{-1}$   
cm$^{-3}$, where  
$B_0$ is the preshock tangential field (Chevalier  
1999). For shock velocity of 80 km s$^{-1}$, preshock density  
of 5 cm$^{-3}$, and $B_0 = 10^{-5}$ G, $n_m$ is equal to 70 cm$^{-3}$ which   
is just consistent with the upper limit derived from [\,S II\,] lines.   
The preshock density of 5 cm$^{-3}$ is at the lower end of the  
typical density range of 5--25 cm$^{-3}$ encountered in interclump media  
of galactic molecular clouds (Chevalier 1999), through which the blast wave  
propagates. This uncertain estimate depends   
strongly on the unknown value of $B_0$, and a possible contamination of the   
optical and infrared emission by an H II region means that   
the preshock interclump density might be higher than 5 cm$^{-3}$. A fiducial   
interclump density of 10 cm$^{-3}$ (Chevalier 1999) might be consistent with  
observations.  
  
The postshock cooling gas should be recombining into neutral hydrogen, which  
can be observed in emission and absorption in the 21 cm H\,I line. Based on the  
preshock density of 10 cm$^{-3}$ and the remnant's size, it is possible to   
estimate the total mass of the neutral hydrogen in the swept radiative shell.  
The bright hemispherical northern part of W28, where most of the optical  
emission is located, is about 30$'$ in diameter (9.5 pc in radius at  
distance of 1.8 kpc), while the southern part is much   
fainter and about twice as large in the east-west direction. This suggests  
that the blast wave broke out of the molecular cloud in the south. For the   
bright  
northern part we estimate that about 700 $M_{\odot}$ should be present in  
the neutral shell. Radio observations in the 21 cm line suggest that there is  
neutral hydrogen associated with W28.  
From a rather sparse cross-pattern  
mapping, Koo \& Heiles (1991) found an excess emission over the galactic  
background at the location of W28, some of it with radial velocities in  
excess of 100 km s$^{-1}$ with respect to the systemic velocity of W28.  
Venger et al. (1982) reported a detection of an expanding H I envelope 82 pc  
in diameter, with mass of $6.9 \times 10^4$ $M_{\odot}$ and expansion velocity  
of 20 km s$^{-1}$, centered on W28. While they identified this massive   
envelope with the expanding shell of W28, this interpretation is not  
consistent with information obtained at other wavelength bands.  
The emission  and absorption in the 21 cm  
line  in this region of the sky is apparently complex, and further  
studies of W28  and its neighborhood in the 21 cm line are warranted.  
Such studies could provide us with crucial information about the  
properties of the radiative shell.  
  
A large amount of ionized gas is seen at optical wavelengths. Long et al.  
(1991) estimated the total H$\alpha$ flux at $1.3 \times 10^{-8}$  
ergs cm$^{-2}$ s$^{-1}$, which at $10^4$ K temperature, $\sim 70$ cm$^{-3}$  
density, and 1.8 kpc distance implies the presence of 300 $M_{\odot}$ of  
ionized gas. This is comparable to the estimated mass of the radiative  
shell. For the preshock density of 10 cm$^{-3}$, blast wave velocity of   
$\sim 100$ km s$^{-1}$ and radius $R$ of 9.5 pc, we estimate that the ionizing  
radiation generated by the radiative blast wave falls short of producing   
the ionizing flux necessary to account for the observed H$\alpha$ flux  
by at least one order of magnitude. An unknown additional source of ionizing  
radiation must be present within the remnant. Possible sources of ionizing  
radiation include evaporating clumps in the remnant's interior (Long et al.  
1991) or gas cooling from X-ray emitting temperatures (e. g., Cox et al. 1999).  
  
The remnant's age is approximately equal to $0.3 R/v_s = 3.6 \times 10^4$ yr,  
with $R = 9.5$ pc and $v_s = 80$ km s$^{-1}$, and assuming that  
the remnant is in the snow-plow stage of the evolution. The estimated  
kinetic energy $E$ of the explosion is $4 \times 10^{50}$ ergs (using   
equation  
[26] of Chevalier 1974). This might be an underestimate because the blast  
wave most likely broke out of the molecular cloud in the south. We also note  
that the derived value of $E$ depends strongly on the assumed distance $d$ to  
the remnant ($E \propto d^{3.12}$, see \S~1 for discussion on the uncertainty   
of the distance). From simple analytical estimates (e. g.,  
see Blondin et al. 1998), the transition from the adiabatic (Sedov) stage to  
the radiative stage should have occurred at   
$t_{tr} \approx 8.6 \times 10^3 E_{51}^{4/17} n_1^{-9/17}$ yr, where   
$E_{51}$ is the explosion energy in $10^{51}$ ergs, and preshock density  
$n_1$ is measured in units of 10 cm$^{-3}$. The radius and velocity of the  
shock at this time were equal to $7.4 E_{51}^{5/17} n_1^{-7/17}$ pc and  
$340 E_{51}^{1/17} n_1^{2/17}$ km s$^{-1}$. While cooling is important at  
$t_{tr}$, the cool radiative shell forms at about  
$1.5 t_{tr}$ (see Table 1 in Blondin et al. 1998), when the shock velocity  
dropped to $220 E_{51}^{1/17} n_1^{2/17}$ km s$^{-1}$.  
  
Significant pressure variations must be   
present within the remnant. For $n_0 = 10$ cm$^{-3}$ and   
$v_s = 80$ km s$^{-1}$, the average pressure $p/k$ in the radiative shell is  
equal to $1.1 \times 10^7$ cm$^{-3}$ K, while Frail \& Mitchell (1998)  
obtain  $p/k \sim 4.6 \times 10^7$ cm$^{-3}$ K   
from observations of the Zeeman splitting in the OH 1720 MHz line.   
At two locations within W28, Claussen et al.  
(1999) detected an even stronger magnetic field, which would imply pressures  
$p/k \sim 4.6 \times 10^9$ cm$^{-3}$ K. Such high pressures are expected  
to be generated by the impact of the radiative shell with the molecular  
clumps (Chevalier 1999). Pressures should be lower than average in the   
southern and the southwestern parts of W28, but no pressure estimates   
are available in this faint breakout region from radio, optical, or   
infrared observations.  
  
\subsection{Interpretation of X-Ray Observations}  
  
The southwestern region has a shell-like morphology, with a fairly good   
correlation between the X-ray and radio emission, and optical emission  
from radiative shocks is not present. This suggests that perhaps description of  
the SW shell spectra in terms of plane-parallel shock and Sedov   
models might be appropriate. This is unlike the central region, where  
X-ray emission is of unknown origin and its center-filled morphology is   
different from the Sedov solution,  while the northeastern shell is located   
in the radiative section of the remnant, close to dense  
molecular clouds with OH masers.   
  
We use our spectral fits (Table \ref{w28ascafit}) to estimate  
physical parameters for the SW  
region. 
The temperature is equal to 1.5 keV, implying a shock velocity of  
1130 km s$^{-1}$. Electron density may be derived from emission   
measure in Table 2, equal to $3.5 \times 10^{56} d_{1.8}^2$ cm$^{-3}$, and  
from the estimated volume $V$ of the X-ray emitting material. The size of the  
shell seen in this location is 8\arcmin\ $\times$ 16\arcmin, or   
$14 \times 9$ pc at  
1.8 kpc distance. Assuming a comparable (14 pc) depth of the shell along the   
line of sight, we arrive at $V \sim 9 \times 10^{57} d_{1.8}^3$ cm$^3$.   
Then, $n_e  = 0.22 f^{-1/2} d_{1.8kpc}^{-1/2}$ cm$^{-3}$,  
where f is a filling fraction of the X-ray emitting material within  
volume $V$.  
With the electron density of 0.22 cm$^{-3}$ and  
ionization age of $1.5 \times 10^{11}$ cm$^{-3}$ s, the shock is about  
$2.2 \times 10^4$ yr old. This is consistent with the estimated remnant's age   
of $3.6 \times 10^4$ yr. The preshock hydrogen density  
$n_0$ is equal to 0.05 cm$^{-3}$, where  
$n_0 = n_e/4.8$ for cosmic abundance plasma and strong shock jump  
conditions. This density is 2 orders of magnitude lower than the  
interclump density in the northern radiative section of the remnant,   
consistent with the observed breakout morphology of W28. The thermal pressure  
$p/k$ of the X-ray emitting plasma, equal to   
$\sim 5 \times 10^6$ cm$^{-3}$ K, is as expected somewhat lower than   
pressures encountered in the bright northern hemisphere. One   
problem is the high shock speed of 1100 km s$^{-1}$,   
because with this speed the shock would propagate to a distance of 40 pc  
in $3.6 \times 10^4$ yr, which is much larger than the remnant's radius.   
In order to investigate this issue, we now discuss the SW shell in the  
framework of spherical (Sedov) shock models instead of plane shocks.  
  
Our Sedov spectral model fit yielded $T_s = 1.2$ keV, and  
ionization timescale of $3.6 \times 10^{11}$ cm$^{-3}$ s. The shock  
temperature in Sedov model is defined as the temperature immediately behind the  
shock. These spectral parameters are consistent with the plane-parallel shock   
model fit parameters; for ion-electron  
equipartition, we expect $T_{shock} = 1.27T_{Sedov}$   
(Borkowski, Lyerly, \& Reynolds 2001).  The ionization timescale   
$\tau_{Sedov}$ is defined as the product of  
the electron density immediately behind the shock and the remnant's age,  
but the mean ionization timescale  
of the gas is equal to 0.202 $\tau_{Sedov}$, which   
should be equal to half of shock age, i.e., $\tau_{shock}/2$. Therefore,   
$\tau_{shock} \approx 0.404\tau_{Sedov}$, which is consistent with   
results obtained from shock and Sedov spectral fits. The physical parameters   
of the remnant can be estimated from the Sedov spectral fit  
parameters.  Using   
$T_s = 1.2$ keV and  $n_et = 3.6 \times 10^{11}$ cm$^{-3}$ s in equations of   
Hamilton, Chevalier, \& Sarazin (1983), we get shock velocity of   
1000 km s$^{-1}$, age of 5100 yr, explosion energy of  
$2.2 \times 10^{51}$ ergs s$^{-1}$, and preshock density of 0.5 cm$^{-3}$.  
The remnant's age estimated in this way is much lower than   
$3.6 \times 10^4$ yr, which is not surprising in view of the problem  
that the shock velocity is too high for the radius of W28,  
encountered with the plane shock analysis.  
There is an order of magnitude discrepancy between the preshock  
densities derived from the Sedov model and estimated from   
the emission measure. This discrepancy, coupled with the low estimated age,  
suggests that Sedov analysis probably does not apply to the breakout  
region of W28, for example, because of large density gradients expected  
there. It is then doubtful that the current shock velocity is as high as  
1000 km s$^{-1}$. For example, material in the southwestern  
shell section could have been shocked long time  
ago, and then adiabatically expanded into the low density region on the  
periphery of the molecular cloud complex. In this case, the blast wave itself   
would have propagated in the low density medium ahead of the observed shell,  
and could be too faint to be detectable with ROSAT or ASCA.   
Observations with a higher spectral  
and spatial resolution, and with a higher S/N ratio are needed in order to   
resolve this issue. The absorbed flux (0.3-10 keV) from this shell section is   
4.5($\pm$0.4)$\times 10^{-12}$ ergs  
s$^{-1}$ cm$^{-2}$, and unabsorbed flux is 4.0($\pm$0.4)$\times  
10^{-11}$ ergs s$^{-1}$ cm$^{-2}$, which is equivalent to X-ray  
luminosity of L$_{\rm x}$ of  1.45$\pm 0.14 \times  10^{34}$ ergs  
s$^{-1}$ at 1.8 kpc distance.  
  
For central spectra, we estimated electron densities  
and filling factors using emission measures. The lower temperature  
component with 6.9$\times$10$^6$ K (0.59 keV) has a density of $0.48 f_{cold}^{-1/2}$   
cm$^{-3}$ and the higher temperature component with 2.1$\times$10$^7$ K  
(1.8 keV) has a density of  $0.19 f_{hot}^{-1/2}$ cm$^{-3}$.   
Assuming that the hot and cold gases   
are in pressure equilibrium, and that the whole volume is filled with the  
emitting gas (i.e., $f_{cold}+f_{hot}=1$), the filling factors of   
cold and hot gases are 0.4 and 0.6, respectively. Electron  
densities in low and high temperature components are 0.75 and 0.25 cm$^{-3}$.  
The thermal pressure $p/k$ of $9.5 \times 10^6$ cm$^{-3}$ K is of the same   
order as our estimates of the pressure in the radiative shell, but   
significantly less than in  
dense regions with OH masers.   
Using the best-fit two-temperature NEI model, we estimated the flux and   
luminosity of X-rays.  
The absorbed flux for the central region (using 2T vnei models) is   
1.7$\pm0.3$$\times 10^{-11}$  
 ergs s$^{-1}$ cm$^{-2}$,  
and unabsorbed flux is 8($\pm$2)$\times 10^{-11}$ ergs s$^{-1}$ cm$^{-2}$, which  
is equivalent to X-ray luminosity of L$_{\rm x}$ of  
2.9($\pm0.7$)$\times 10^{34}$ ergs s$^{-1}$, which is twice as high as that   
of the SW region. The northeastern region spectral fit gives an X-ray   
luminosity similar to that of central region.  
  
The limb brightened northeast region (the left ``ear'') has a shell-like  
appearance and it appears to be spatially coincident with radiative shell   
as indicated by a very good spatial match with bright radio and optical  
filaments (Dubner et al. 2000; see also Fig.~\ref{w28sb}). The temperature of   
the X-ray emitting gas is equal to 0.56 keV (6.5$\times 10^6$ K).   
This temperature is   
significantly higher than expected from a spherically symmetric remnant  
at the late radiative stage of the evolution with parameters as estimated  
above. For example, the postshock temperature at $t_{tr}$ is equal to  
$0.14 E_{51}^{1/17} n_1^{2/17}$ keV in the Sedov model. Falle (1981)  
demonstrated that the emission-measure weighted temperature of the  
X-ray emitting gas at times longer than $t_{tr}$ is no longer related to the  
shock speed. It fluctuates instead by about 50\%\ around the value just  
quoted until at least $10 t_{tr}$ (see Fig. 8 in Falle 1981). This  
discrepancy between the expected and the observed temperature cannot be  
solved by simply postulating a much higher interclump density, because our  
density estimates based on the emission measure rule out very high   
densities of the X-ray emitting gas. Note however that it might be difficult  
to detect gas with very low temperatures because of the high   
ISM absorption along the line of sight to W28.   
  
Our best estimate of the total X-ray emitting mass is   
$\sim 20-25 d_{1.8}^{5/2} M_{\odot}$:  
most of it in the center and 8 $M_{\odot}$ in the northeastern and  
southwestern shells. This is less than expected from a radiative remnant  
with parameters appropriate for W28. Numerical calculations of   
Hellsten \& Sommer-Larsen (1995) indicate that about 10\%\ of the total   
shocked gas (by mass) is emitting in X-rays at $3.6 \times 10^4$ yr, for  
a model SNR remnant with $E_{51}=1$ and $n_1=1$. We would then expect   
$\sim 100$ $M_{\odot}$ of X-ray emitting gas in W28, mostly at low  
(0.14 keV) temperatures. This is consistent with an analytical estimate  
of 120 $M_{\odot}$, using equation (33) of Cui \& Cox (1992).  
The observed X-ray emission comes from a much smaller  
amount of hotter gas, suggesting a somewhat different SNR evolution than  
implied by standard spherically-symmetric SNR models in a uniform ambient   
medium.   
This is ``fossil'' radiation, a conclusion supported by long cooling   
timescales of the gas.  
We estimated cooling timescales using temperatures and densities derived  
above; for isobaric cooling this timescale is   
$6 \times 10^{6}$ yr in the SW, and $3.7 \times 10^{5}$ yr for northeastern   
shell gas. This is much longer than the age of the remnant, so that the  
X-ray emitting gas cannot be associated with    
fast radiative shocks or with gas cooling from high temperatures.  
Instead, this material was shocked much earlier   
in the evolution of the remnant, and its temperature and density may have  
been modified since then by processes such as evaporation, electron thermal  
conduction, and mixing induced by various hydrodynamical instabilities.   
We consider next whether the evaporation model of White \& Long (1991) or  
the conduction model of Cox et al. (1999) can explain the observed X-ray   
properties of W28.  
  
The evaporation model of White \& Long (1991) is in conflict with    
ASCA spectra. In this model, because of the short timescale   
of evaporative flows induced by saturated thermal conduction, heavy-element  
ions in gas evaporated from small clouds are strongly underionized with respect  
to CEI plasmas when they are mixed with the  
hot diffuse gas occupying most of the SNR interior. Other processes such as  
hydrodynamical instabilities would further hasten mixing with the hot gas, so  
that ionization should occur mostly in the hot diffuse gas. Long et al. (1991)   
obtained an age of only 2400 yr for W28 using the evaporation model, which in   
combination with electron densities estimated from  
emission measures implies the presence of strongly underionized  
gas in W28. But ASCA spectra reveal gas nearly in ionization equilibrium,  
with ionization timescales at least one or two orders of magnitude longer  
than expected. X-ray spectra alone are sufficient   
to reject evaporation model for W28. But perhaps a more convincing argument  
against this model is provided by the presence of radiative shocks in the   
northern half of the remnant, which are not expected in the nonradiative  
evaporative model considered by Long et al. (1991). The failure of this  
model for W28 further supports conclusions of Chevalier (1999)   
who argued against the White \& Long (1991) model as the explanation for the   
MM SNRs.  
  
The radiative shell model of Cox et al. (1999) and Shelton et al. (1999), in  
which electron thermal conduction modifies the interior of an SNR evolving in  
a uniform ambient medium, provides reasonable values for the central density  
and temperature. Using equations of (3) and (10) of Cox et al. (1999), with  
$E_{51} =1$, $n_1 = 1$, and the Spitzer's conductivity coefficient, we arrive  
at the central electron density of 0.93 cm$^{-3}$ at the present   
($3.6 \times 10^4$ yr) time. The central temperature of 0.52 keV is then  
obtained by  
equating the central pressure to the pressure within the radiative shell.  
There is obviously a good agreement with the densities and temperatures  
of the cool temperature component found in the central regions of W28. But   
the conduction model predicts an X-ray emitting mass which is even larger   
than in   
the standard radiative shell model without conduction considered by  
us previously, in apparent disagreement with observations. In addition, the   
surface brightness of W28 is too sharply   
peaked at the center in comparison with the conduction model (it is sharper   
than that of W44; see Rho \& Petre 1998), and the clumpiness seen in X-ray   
images is in contrast to a smooth gas distribution expected in this   
model. But the most serious challenge for  
any models in which conduction plays a major role might be the presence of  
temperature gradients in W28. The observed temperature differences must  
have originated early in the remnant's evolution, before the radiative shell  
formed, when efficient conduction   
would have obliterated temperature gradients in the remnant's interior.  
Using a self-similar conduction model of Cui \& Cox (1992), we find that at the  
time of the shell formation at $1.3 \times 10^4 $ yr, temperature dropped  
from its central value by a factor of 2 at a sphere enclosing the innermost  
75 $M_\odot$ of X-ray emitting gas. In  
apparent conflict with the model predictions, we observe larger  
temperature differences within much smaller amounts of X-ray emitting gas.   
Efficient thermal conduction might be inconsistent with the observed   
X-ray properties of W28.  
  
The X-ray properties of W28, in particular the simultaneous presence of hot   
(1.5 keV) and cold (0.5 keV) components with long ionization timescales,   
are in conflict with predictions of the evaporation model of White \& Long   
(1991) and perhaps are inconsistent with predictions of the  
radiative shell model with conduction (Cox et al. 1999). Apparently   
evaporation is less important than postulated by White \& Long (1991), and  
conduction might be less efficient than postulated in both models.  
The radiative shell interacting with clumpy molecular clouds, as  
considered by Chevalier (1999), seems to provide a suitable framework for  
W28. In this model, X-ray emission at the center is a ``fossil'' radiation,   
whose properties   
are not fully understood at this time. Uniform temperature gradients   
seen in MM SNRs led Chevalier (1999) to argue that conduction  
is generally important in their interiors. This argument cannot be used  
for W28 because of the presence of temperature variations, including  
a large-scale temperature gradient. This temperature gradient may   
be attributed to different preshock densities (and hence shock speeds) in the   
northeast and southwest, due to interaction with molecular clouds in the north and  
northeast, and the breakout in southwest. In absence of efficient conduction  
on spatial scales comparable to the radius of the remnant, appreciable   
temperature gradients of the opposite sign to density gradients are generally  
expected. Their presence in W28 suggests that conduction might have been at   
least partially inhibited by magnetic fields, perhaps preferentially in the  
direction perpendicular to the field lines of large-scale magnetic fields.   
Another possibility involves poorly understood mass and energy exchange  
processes between the hot and tenuous interclump medium and dense cool clumps  
of gas embedded in this medium. While idealized models dominated by either  
evaporation or by conduction failed to explain observations, conduction,  
evaporation and mixing induced by various hydrodynamical   
instabilities must still be present in view of the clumpy nature of molecular  
clouds. Their relative importance remains to be determined by  
comparing future model calculations with X-ray spectra and morphology of W28.   
W28 should provide a stringent test for future modeling because of the  
presence of the appreciable temperature variations.

\subsection{Comparison of W28 and other mixed-morphology SNRs}  
  
Understanding of MM SNRs requires a careful assessment of their X-ray   
properties, so that competing models can be tested on the basis of their   
ability to explain these properties. For example, uniform temperature profiles   
characteristic of MM remnants are well explained by models with efficient  
thermal conduction, but the presence of large temperature gradients in even   
a small subset of these SNRs might pose a problem for these models. Because    
X-ray spectra revealed multitemperature plasma in W28,   
we contrast X-ray properties of W28 with other MM SNRs.  
  
W28 is known to be an archetype MM SNR along with W44 (Long et al. 1991;  
Rho \& Petre 1998), because both show typical mixed morphology  
with center-filled X-rays and the shell-like radio morphology. However,   
our images and spectral analysis show that X-ray properties of W28 are   
somewhat different. First, unlike in other known MM SNRs, a high  
temperature plasma is certainly present, as evidenced by the Fe K$\alpha$ line   
and the thermal continuum extending up to 10 keV. The southwestern shell has a   
high temperature of 1.5 keV, and a high temperature component of 1.8 keV is  
also present at the center.   
  
X-ray emitting gas in W44 has temperature of 0.4--0.9 keV without any  
hard  emission above 6 keV present (Harrus et al. 1997; Rho et al.  
1994). Other MM  SNRs (e.g., see Rho \& Petre 1998) such as MSH 11-61A  
(Rho  1995), HB21 (Lee et al. 2001), 3C391 (Rho \& Petre 1996; Chen \& Slane  
2001), and 3C 400.2 (Saken et al. 1995; Yoshita et al. 2001), have gas in the  
temperature range of 0.4--0.9 keV.    
ASCA data of HB21 also imply a low temperature of 0.6 keV (Lee et al.  
2001), similar to results based on ROSAT data (Rho \& Petre 1998).  
Plasma temperatures in these MM SNRs are between 0.4--0.9 keV. Other  
remnants also have gas in this range of temperatures (Rho \& Petre  
1998). In contrast, there is gas with a much higher temperature of 1.5  
keV in the southwestern shell and 1.8 keV in the central region of  
W28.

The second difference is the presence of spectral variations in W28,   
in contrast to other MM SNRs which have  
uniform temperatures as revealed by spectral mapping  
(Rho et al. 1994; Rho \& Petre 1998).  
The spectrum of the southwestern region   
is different  
from the center. The southwestern spectrum  
seems to show a high (1.5 keV) temperature  
component alone without the soft component.  
This may be partially accounted for by a higher absorption along the line of  
sight in the SW, but the poor S/N spectrum of the SW region ultimately limits  
our ability to understand the origin of these spectral differences.  
The spectrum of the northeastern region is also different from  
the center. The northeastern spectrum  
seems to show a low (0.56 keV) temperature  
component alone without requiring a second component.  
The central spectrum  
is noticeably different from the southwestern and northeastern spectra.    
Our results show the presence of pronounced spectral  
variations within W28, quite unlike what is seen in more  
typical MM SNRs.  
Another difference is that X-ray mass of W28 is much  
less than those of other MM SNRs such as W44 and 3C391 (Rho et al. 1994;  
Rho \& Petre 1996).

W28 has still common characteristics with other MM SNRs; X-ray  
emission is thermal  with rich line emission as shown by the ASCA  
spectra, and the spectral fitting implies that the abundances are close to  
solar (ISM) abundances, without any evidence for ejecta material. There is  
also the prominent central X-ray emission within the well-defined radio shell.  
Therefore, W28 is still an MM SNR, but   
with a number of different X-ray properties;  
1) partial shells are present, 2) X-ray emission is extremely clumpy,  
particularly in the northeastern shell and at the center,  
3) temperature is higher than in other MM SNRs, and   
4) spectral variations are present due to  
temperature differences. The difference between the  
northeastern and southwestern shells appears to be related to differences in  
preshock densities, while both hot and cool components are present in the  
central region.  
  
Recently discovered MM SNRs  
such as G359.1-0.5 (Bamba et al. 2000),  
Sgr A East (Maeda et al. 2001) and Kes 27 (Rho et al. 1998)   
also share some of these characteristics  
with W28. They show evidence for very hot temperatures  (1.5 -- 4 keV),  
and mixed morphology between X-rays and radio, and they are located    
in dense environments, possibly interacting with nearby clouds.  
Several of them even show evidence of SN ejecta,   
which strongly suggests that they are younger than more well-known  
MM SNRs W44 and 3C 391.   
  
\section*{Summary}  
  
1. The overall shape of X-ray emission in W28 is elliptical, with the  
emission concentrated at the center of the remnant. The interior emission  
is sharply peaked at the center and its morphology is highly patchy.  
Partial shells are present in the northeast and southwest, and the  
northeastern shell also has a patchy, partially broken appearance.   
There are ear-like segments of limb-brightened shells in the northeast  
and northwest. The X-ray surface brightness between the two ear-like  
patches stays very flat from  east to west along Dec -23$^{\circ}$  
15$'$. The surface brightness is composed of 3 components: 1) the  
innermost central region with centrally peaked brightness, 2) smooth  
component, and 3)  northeastern and  southwestern shell  components.  
The observed surface brightness distribution in the SNR interior  
implies a radial density profile  described by $n/n_c = 1 - r/R$, where  
$n(r)$ is plasma density as a function of distance $r$ from the  
remnant's center, $n_c$ is the central density,  and $R$ is the  
remnant's radius. This profile is steeper than profiles usually found  
in other mixed-morphology SNRs. The overall shape of the remnant is far  
from uniform and circular, typical for supernova remnants; it is dented  
in the north and east because of the presence of  molecular clouds.

2.   
The ASCA spectra show emission lines of  
Ne, Mg, Si, and Fe K$\alpha$, and continuum at least up to 7 keV.  
ASCA/ROSAT spectra reveal spectral variations across the remnant:  
while the southwestern and northeastern shells are fit well by  
a one-temperature model, the central emission requires a two-temperature  
model with hot and cold components.  The southwestern shell can be fit well  
by the plane shock model with a temperature of 1.5 keV,  
an ionization timescale of $1.5 \times 10^{11}$ cm$^{-3}$ s,    
and a line-of-sight absorption of   
$1.2 \times 10^{22}$ cm$^{-2}$. The  
northeastern shell has a lower temperature of $0.56$ keV,  
a longer ionization timescale of $1.7 \times 10^{13}$  
cm$^{-3}$ s, and $N_H = 0.8 \times 10^{22}$ cm$^{-2}$.  Unlike for the  
southwestern and northeastern shells, the central emission cannot be  
fit well by a single temperature model. Two components are required,  
a hot temperature component with  
$kT_{h} = 1.8 \pm0.5$ keV and a long ionization timescale  
of $5 \times 10^{11}$ cm$^{-3}$ s, and a low  
temperature component with $kT_{l} = 0.6$ keV, $n_et = 2 \times 10^{12}$  
cm$^{-3}$ s. Note that with the possible exception of the SW shell, the long  
derived ionization timescales imply that the gas is close to the ionization   
equilibrium.  
The X-ray luminosity of W28 is $6 \times 10^{34}$ ergs s$^{-1}$  
(this includes only regions observed by ASCA/SIS),    
and the estimated total X-ray mass is rather small,   
only $\sim 20 - 25 M_\odot$, much less than expected from standard SNR models.

3. Observations at wavelength bands other than X-rays strongly suggest  
that W28 is in a radiative stage of evolution, at least in the northern half  
of the remnant where the interaction with the molecular clouds has been  
detected. This remnant then belongs to a class of SNRs considered by  
Chevalier (1999), with a radiative shell interacting with clumpy molecular   
clouds, in which X-ray emission in the center is a ``fossil'' radiation.   
The properties of X-ray emitting gas are broadly consistent with this scenario,  
but when compared with standard spherically-symmetric SNR models   
in a uniform ambient  
medium, the gas is hotter, there is less X-ray emitting gas than expected,  
and of course the centrally-filled morphology cannot be explained by such  
models. The material which we are observing at the center  
was likely shocked early  
in the evolution of the remnant, and its temperature and density may have  
been modified since then by processes such as evaporation, electron thermal  
conduction, and mixing induced by various hydrodynamical instabilities.  
But the evaporation model of White \& Long (1991) is in conflict with  
observations, and efficient thermal conduction might not be consistent  
with large temperature variations detected in W28. Both evaporation and  
thermal conduction might be less efficient than assumed in these idealized  
models, while the clumpy nature of molecular clouds plays an uncertain  
but perhaps even a crucial role in determining X-ray morphology and spectra.  
  
4. W28 differs from other known MM SNRs by: 1) the presence of hot (1.5 keV)  
gas, 2) spectral variations are clearly present, the SW shell is much hotter  
than the NE shell, and a mixture of hot and cold gas is present at the  
center, 3) X-ray emission is more   
knotty and clumpy than in other MM SNRs, and partial shell-like structures  
are present. The large scale temperature gradient seems to to be related  
to the ambient ISM density, with the cooler gas associated   
with the molecular clouds and the hotter gas with the fainter   
breakout region.   
The separation between X-ray emitting gas and radiative shocks emitting in  
infrared and optical is smaller than in other MM SNRs, more  
similar to shell-like SNRs.    
   
5. We report a marginal excess of X-ray emission in the ROSAT image   
at the position of   
the  pulsar, PSR B1758-23,  which is located at the northern  boundary  
of W28. The count rate implies a luminosity of $1.2 \times 10^{32}$ ergs  
s$^{-1}$, assuming  a line-of-sight absorption of $8 \times 10^{21}$  
cm$^{-2}$ and a power law spectrum with $\Gamma = 2$ (equal to the  
spectral index of  the  Crab Nebula). In addition to the pulsar, two  
isolated X-ray sources appear a few arcminutes away from the  
eastern boundary. The northern source is a point source, likely a  
foreground star with no counterpart at other wavelengths. The  
southern source which appears to be extended, can be identified  
with an HII region.

\acknowledgements  
We thank Robert Petre for helpful discussion about  the earlier version  
of the paper, and Ho-Gyu Lee for help in generating ASCA mosaiced images.   
We also acknowledge very helpful comments and suggestions by an anonymous   
referee.  
This research has made use of data obtained from the High Energy   
Astrophysics Science Archive  
Research Center (HEASARC), provided by NASA's Goddard Space Flight Center.   
J. R. acknowledge the support  
of the Jet Propulsion Laboratory, California Institute of Technology,  
which is operated under contract with NASA.

\def\Mdot{\dot M}

\begin{table}  
\caption{ROSAT and ASCA Observations of W28 (G6.4-0.1).}  
\label{w28obser}  
\begin{center}  
\begin{tabular}{lccccccc}  
\\  
\hline \hline  
 &pointing &Sequence No. &observation& exposure &\multicolumn{2}{c}{center} \\  
 &         &       & date      &time (s)     &R.A. &Dec.\\  
\hline  
 ROSAT  \span\omit  \span\omit  \span\omit && \\  
&center&wp500129 &Mar5, 92& 4224 &18:00:49 & -23:22:33  \\  
 &NE& rp500236&Sep9-10, 93&10,476&18:01:12 & -23:25:60 \\  
 &SW& rp500237&Sep 9, 93&9100 &17:59:59 & -23:31:60 \\  
\hline  
ASCA \span\omit  \span\omit  \span\omit &&  \\  
 & NE  & ad51014000  & Mar4, 94 & 28,000 & 18:01:06.00 & -23:13:51.6 \\   
 & center1 & ad51022000 &Mar4-Apr4, 94,&21,000 & 18:00:41.76 & -23:26:36.6 \\   
 & SW     & ad54003000 & Mar31, 96 &11,500 & 17:59:18.0   & -23:45:56.5 \\  
 & center2 & ad54003010 & Mar31, 96 &10,500 & 18:00:22.8 &-23:20:18.0 \\  
\hline  
\end{tabular}  
\end{center}  
\end{table}  
  
\begin{table*}  
\caption{Spectral Fits and Inferred Physical Parameters}  
\label{w28ascafit}  
\begin{center}  
\begin{tabular}{lcccccccccc}  
\tableline \tableline \tableline  
Region     &   
\multicolumn{3}{c|}{Center} &  \multicolumn{2}{c|}{SW shell} & \multicolumn{1}{c|}{NE shell} \\  
\hline  
Model     & A\tablenotemark{a} & B\tablenotemark{b} &C\tablenotemark{c} &D\tablenotemark{d} & E\tablenotemark{e} & F\tablenotemark{f} \\  
\tableline  
$\Delta\chi^2$$^g$ & 1.4 &1.1 & 1.1 &  0.96 & 0.93 & 1.01\\  
$N_H$ ($10^{22}$cm$^{-2}$) &  0.73$^{+0.04}_{-0.06}$  & 0.66$\pm$0.06 & 0.79$\pm$0.15 & 1.28$\pm$0.25 & 1.26$^{+0.24}_{-0.16}$ & 0.77$\pm$0.08 \\  
\hline \hline  
\multicolumn{7}{l}{Low temperature component }  \\  
\hline  
$kT_l$ (keV)  & 0.58$^{+0.04}_{-0.07}$ & 0.59$\pm$0.05 & 0.35$\pm$0.15 & 1.2 ($>$0.8) & 1.5$\pm$0.2 & 0.56$\pm$0.06\\  
$\tau_l$ (10$^{12}$ cm$^{-3}$ s) & -- & 2.0 ($>$1) & 1.8 ($^{+2.7}_{-0.8}$) & 0.36$^{+0.20}_{-0.16}$ & 0.15$^{+0.17}_{-0.09}$  & 17 ($>$7) \\  
Abundance (fixed) &\multicolumn{3}{c}{2/3} & \multicolumn{2}{c}{2/3} & \multicolumn{1}{c}{2/3}\\  
$EM_l/(4\pi d^2)$ (10$^{12}$ cm$^{-5}$)  & 4-5.3 & 3.5-4.6 & 6.3-8.5 &0.8-1 & 0.9-1 & 1.9-2.3\\  
\hline \hline  
\multicolumn{7}{l}{High temperature component }\\  
\hline  
$kT_h$ (keV) (fixed$^h$)  &\multicolumn{3}{c}{1.8($^{+0.7}_{-0.5}$)} &\multicolumn{2}{c}{--}& \multicolumn{1}{c}{--}\\  
$\tau_h$ (10$^{12}$ cm$^{-3}$s) (fixed$^h$)  &  \multicolumn{3}{c}{0.5($^{+0.50}_{-0.43}$)}&\multicolumn{2}{c}{--} & \multicolumn{1}{c}{--}\\  
Fe (fixed) & \multicolumn{3}{c}{1.1($^{+0.9}_{-0.6}$)}  &\multicolumn{2}{c}{--} &\multicolumn{1}{c}{--}\\  
$EM_h/(4\pi d^2)$ (10$^{12}$ cm$^{-5}$)  & \multicolumn{3}{c}{0.65-0.8} &\multicolumn{2}{c}{--} &\multicolumn{1}{c}{--}\\  
\hline \hline  
\multicolumn{2}{l}{ Angular Diameter} & \multicolumn{4}{c}{50$'$$\times$45$'$}\\  
\multicolumn{1}{l}{ Shell Radius} & \multicolumn{3}{c}{--} &  
\multicolumn{2}{c}{28$'$ (14.5$\times$d$_{1.8kpc}$ pc)$^i$}  
& \multicolumn{2}{c}{18$'$ (9.5$\times$d$_{1.8kpc}$ pc)$^j$ } \\  
\hline \hline  
$n_e$ (cm$^{-3}$, from EM) &  
\multicolumn{3}{c}{$0.48(d_{1.8kpc} f)^{-1/2}$, $0.19(d_{1.8kpc} f)^{-1/2}$} &   
\multicolumn{2}{c}{$0.22(d_{1.8kpc} f)^{-1/2}$} & \multicolumn{1}{c}{$0.65(d_{1.8kpc} f)^{-1/2}$ }\\  
Flux (ergs s$^{-1}$ cm$^{-2}$)  &  
\multicolumn{3}{c}{8$\times 10^{-11}$}  &  
 \multicolumn{2}{c}{4$\times 10^{-11}$} & \multicolumn{1}{c}{4.5$\times 10^{-11}$  }\\  
Luminosity (ergs s$^{-1}$) &  
\multicolumn{3}{c}{2.9$\times10^{34}$$d_{1.8kpc}^2$ }  &   
\multicolumn{2}{c}{1.45$\times 10^{34 }$$d_{1.8kpc}^2$ } &  
\multicolumn{1}{c}{1.63$\times 10^{34}$$d_{1.8kpc}^2$}     \\  
\hline \hline  
\tableline  
\end{tabular}  
\end{center}  
\tablenotetext{a}{$mekal+nei$}  
\tablenotetext{b}{$nei+nei$}  
\tablenotetext{c}{$sedov+nei$}  
\tablenotetext{d}{$sedov$}  
\tablenotetext{e}{$pshock$}  
\tablenotetext{f}{$nei$}  
\tablenotetext{g}{$\Delta\chi^2$ is reduced $\chi^2$, and typical degree of freedom is $\sim$650. The errors
of the best fits for each region with 99\% confidence are given.}  
\tablenotetext{h}{$kT_h$  and the Fe abundance were determined from fits to the high energy   
  (4-10 keV) spectra alone, and $\tau_h$ is determined using the Fe K$\alpha$ line centroid.}  
\tablenotetext{i,j}{Measured from the center to the SW and NW shells, respectively.}  
\end{table*}

\clearpage  
  
\begin{figure}  
\caption{The mosaiced ROSAT PSPC image (0.5-2.4 keV) of W28.  
The grey-scale ranges from 0.0008 to 0.029   
counts s$^{-1}$ arcmin$^{-2}$.}  
  
\label{w28pspc}  
\end{figure}  
  
\begin{figure}  
\caption{ASCA GIS fields of view (four pointings)  
superposed on the ROSAT image.}  
\label{w28ascafov}  
\end{figure}  
  
\begin{figure}  
\caption{ASCA greyscale image (0.5-10 keV) and PSPC contours. The grey-scale image  
range is  (1.0 - 5.5) $\times 10^{-5}$ counts s$^{-1}$ arcmin$^{-2}$.}  
\label{w28ascaall}  
\end{figure}

\begin{figure}  
\caption{ Spectral extraction regions (solid circle and ellipses) superposed on the PSPC image.  
Right "ear-like" structure and northern flat boundary are also marked (dashed lines). }  
\label{w28regionmark}  
\end{figure}  
  
\begin{figure}  
\caption{The PSPC greyscale image superposed on  
contours of the 328 MHz radio continuum image (from Dubner et al. 2000): the radio  
morphology is shell-like and X-rays are centrally peaked.}  
\label{w28xrayradio}  
\end{figure}  
  
\begin{figure}  
\caption{Soft ($E < 0.5$ keV) PSPC X-ray image superposed on PSPC total  
brightness contours. The greyscale  
is from 0.0008 to 0.0043 counts s$^{-1}$ arcmin$^{-2}$.}   
\label{w28soft}  
\end{figure}

\begin{figure}  
\caption{The PSPC spectral hardness ratio (0.9--2.2 keV/0.3--0.9 keV) image   
with a greyscale range from 1.1 to 1.4,   
superposed on PSPC total brightness contours.   
The southern half of the  
remnant is harder than the northern half.}  
\label{w28shr}  
\end{figure}  
  
\begin{figure}  
\caption{  
 Hard (3--10 keV) ASCA image superposed on PSPC contours.  
The southwestern shell and center are harder than the northeast.  
The grey-scale image  
range is  (1 - 3) $\times 10^{-5}$ counts s$^{-1}$ arcmin$^{-2}$.}  
\label{w28ascahard}  
\end{figure}

\begin{figure}  
\caption{X-ray ({\it solid line}) and radio ({\it dotted line}) surface   
brightness profiles of W28 in the north (a), southwest (b),  
southeast (c), and northeast (d).  Position angles are marked in each panel.   
The X-ray surface brightness is centrally peaked, but there are also shell  
structures at about 20$'$ from the center in the southwest and  
northeast.  
The dashed line in the northern sector corresponds to a radial density   
profile described in the text. }  
\label{w28sb}  
\end{figure}

\begin{figure}  
\caption{Comparison of W28 spectra for various regions. The best fits
are shown, which are derived using 5 spectra from SIS0, SIS1, GIS2, GIS2
and ROSAT PSPC.
(a) The central ({\it crosses}) and SW ({\it squares}) SIS0 spectra. Soft component  
is not present in the SW.  (b) ASCA SIS spectra of the  
northeastern shell.  The best-fit NEI model with  
$kT_e = 0.56$ keV, $N_H = 0.8 \times 10^{22}$ cm$^{-2}$,  and  
the residuals are  shown.  
(c) The ASCA SIS and ROSAT/PSPC spectra of W28 for the  
central region showing hard emission and the Fe K$\alpha$ line.   
A two-temperature NEI model with $kT_{h} = 1.8\pm0.5$ keV and  
$kT_{l} = 0.6$ keV is also shown.  
(d) Two-temperature model of the central spectrum (only SIS0 spectrum is shown) with the hot and cold   
components indicated (total -- {\it solid line}, cold -- {\it dash-dotted},  
hot -- {\it dotted}).  
}   
\label{w28spec}  
\end{figure}

\begin{figure}  
\caption{ A portion of the PSPC image with the pulsar position marked. Its  
position is at least 30$''$ away from a weak X-ray peak. The ellipse is  
an error box of the pulsar position. The diffuse emission is  northern
part of W28.}  
\label{w28pulsar}  
\end{figure}

\clearpage  
{Fig. 1,2,3 are  jpeg files.} 

  



\psfig{figure=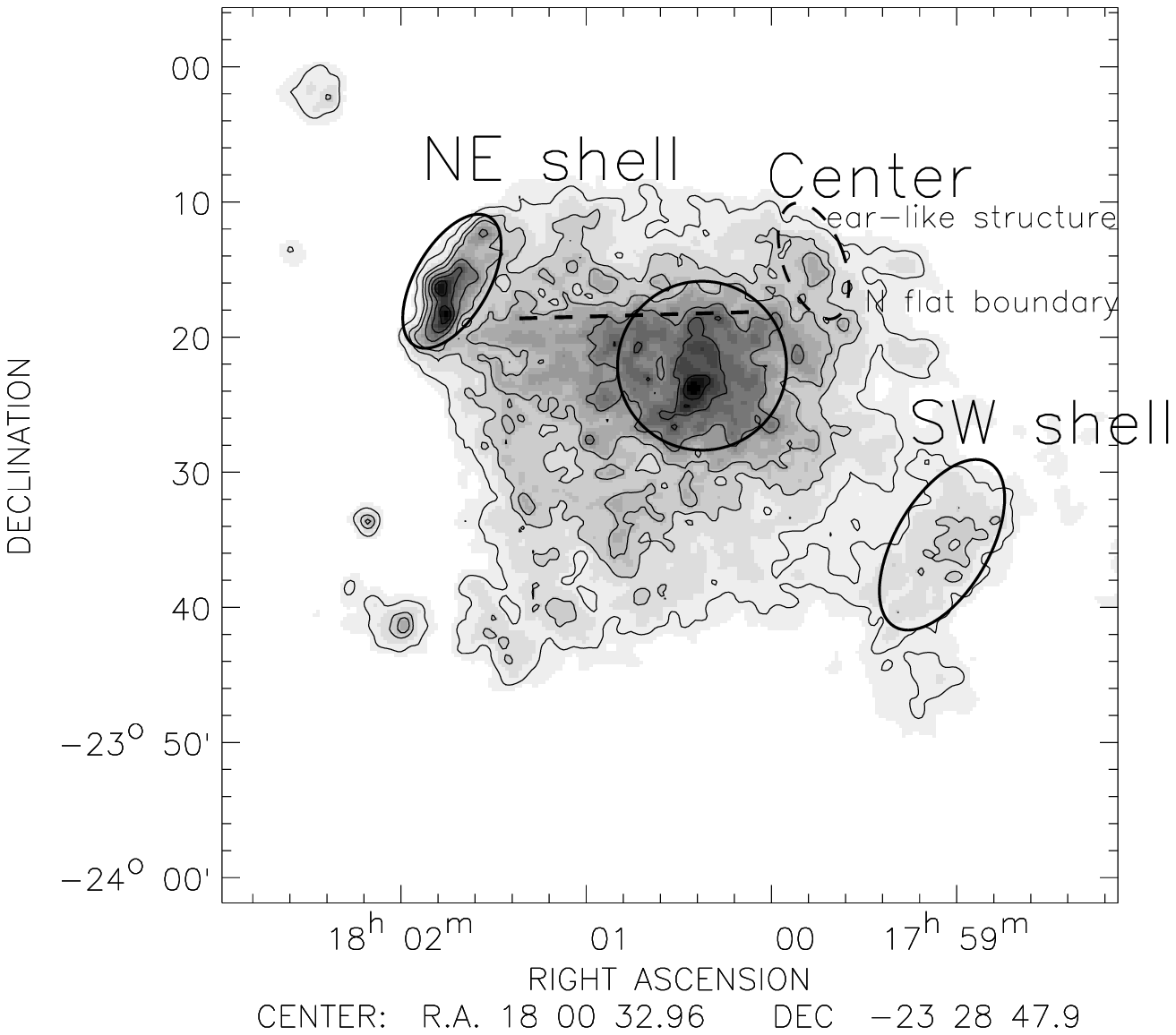}  
\vskip 13truept  
{Fig. 4}  
  
\clearpage  
{Fig. 5,6,7, and 8 are jpg files.}
%
%
%
%
%
%
\clearpage  
  
\vbox{  
  
\centerline{\hbox{  
\psfig{figure=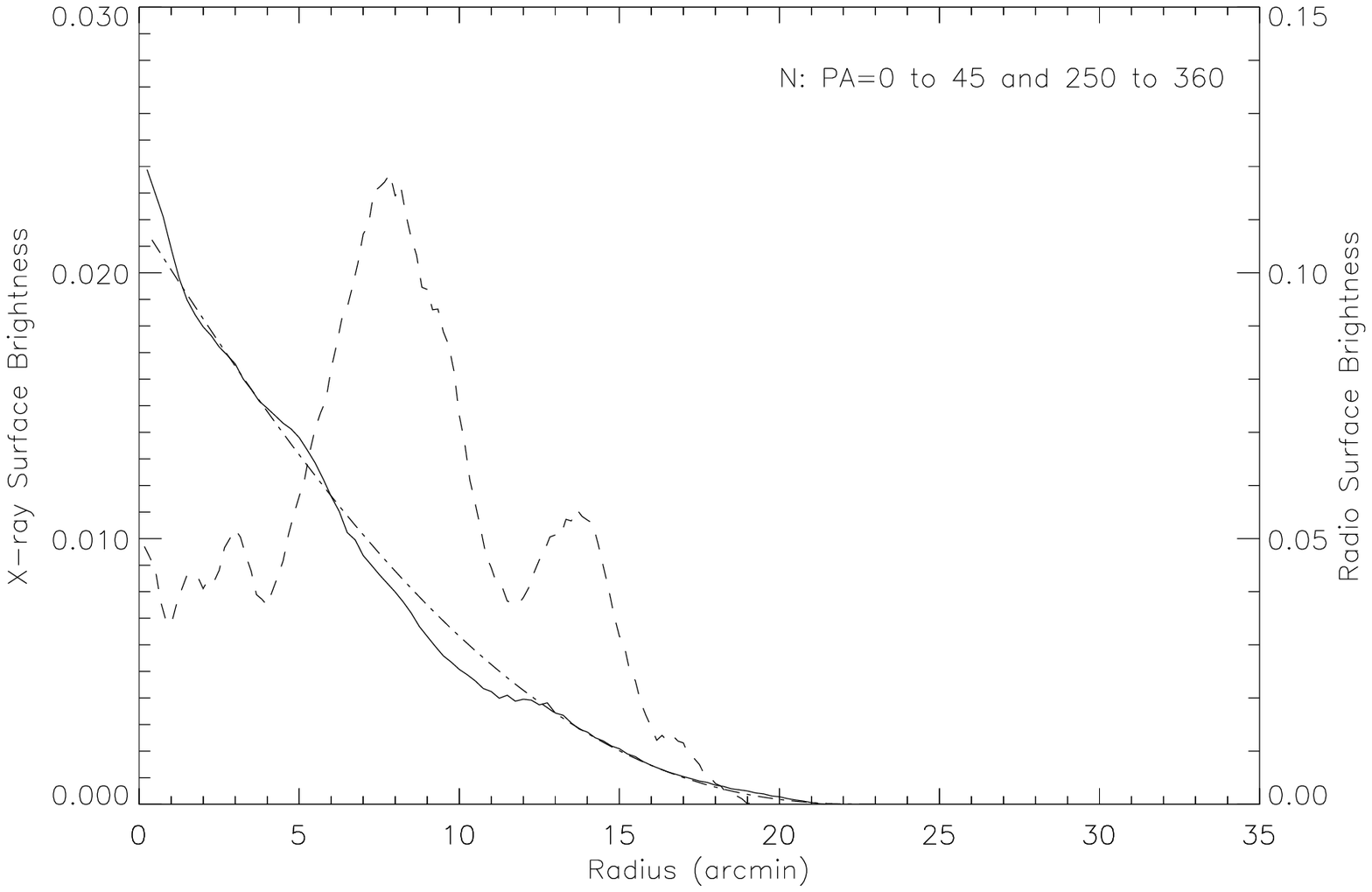,width=8truecm,height=7truecm}  
\psfig{figure=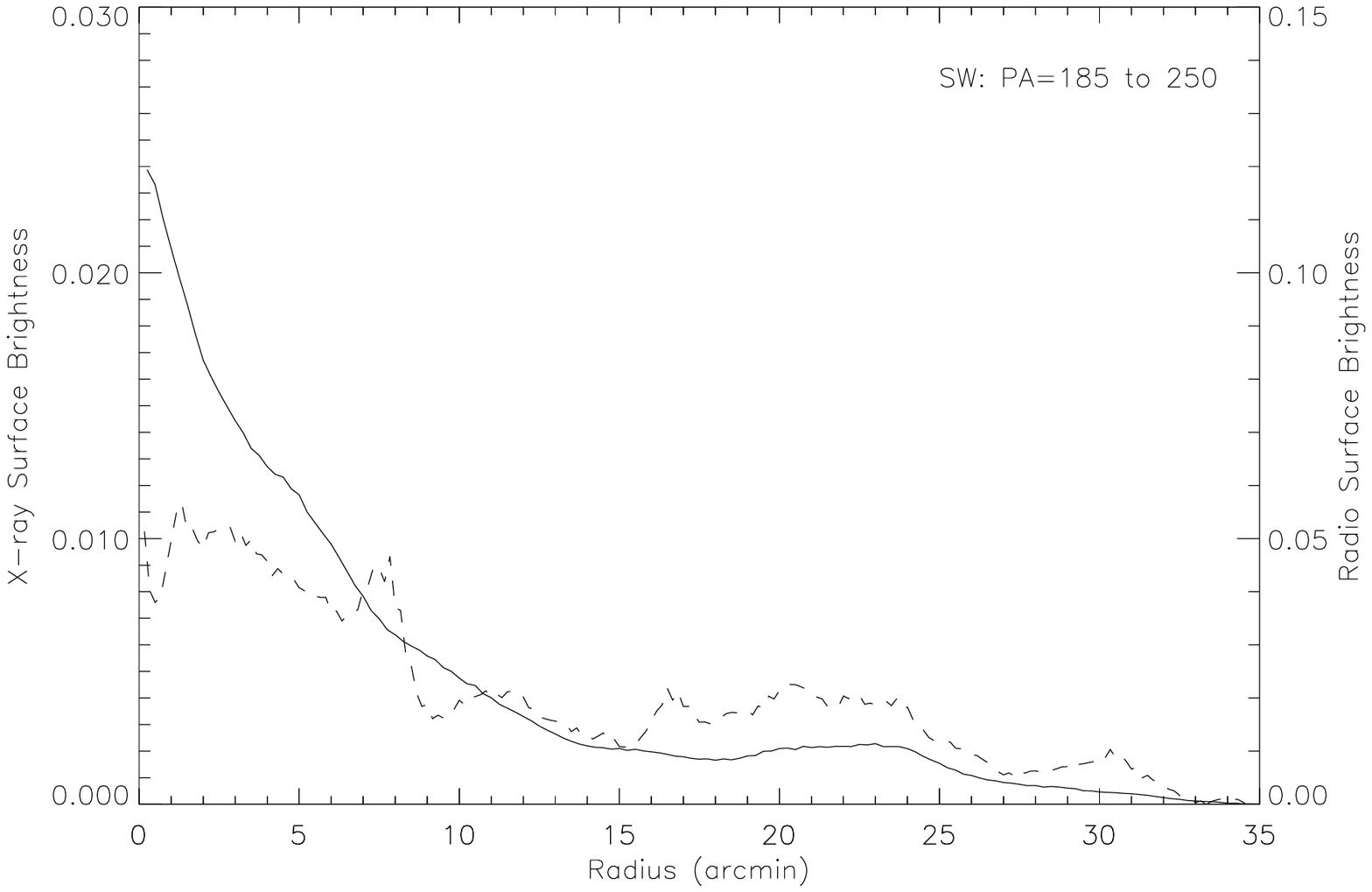,width=8truecm,height=7truecm}  
}}  
\centerline{\hbox{  
\psfig{figure=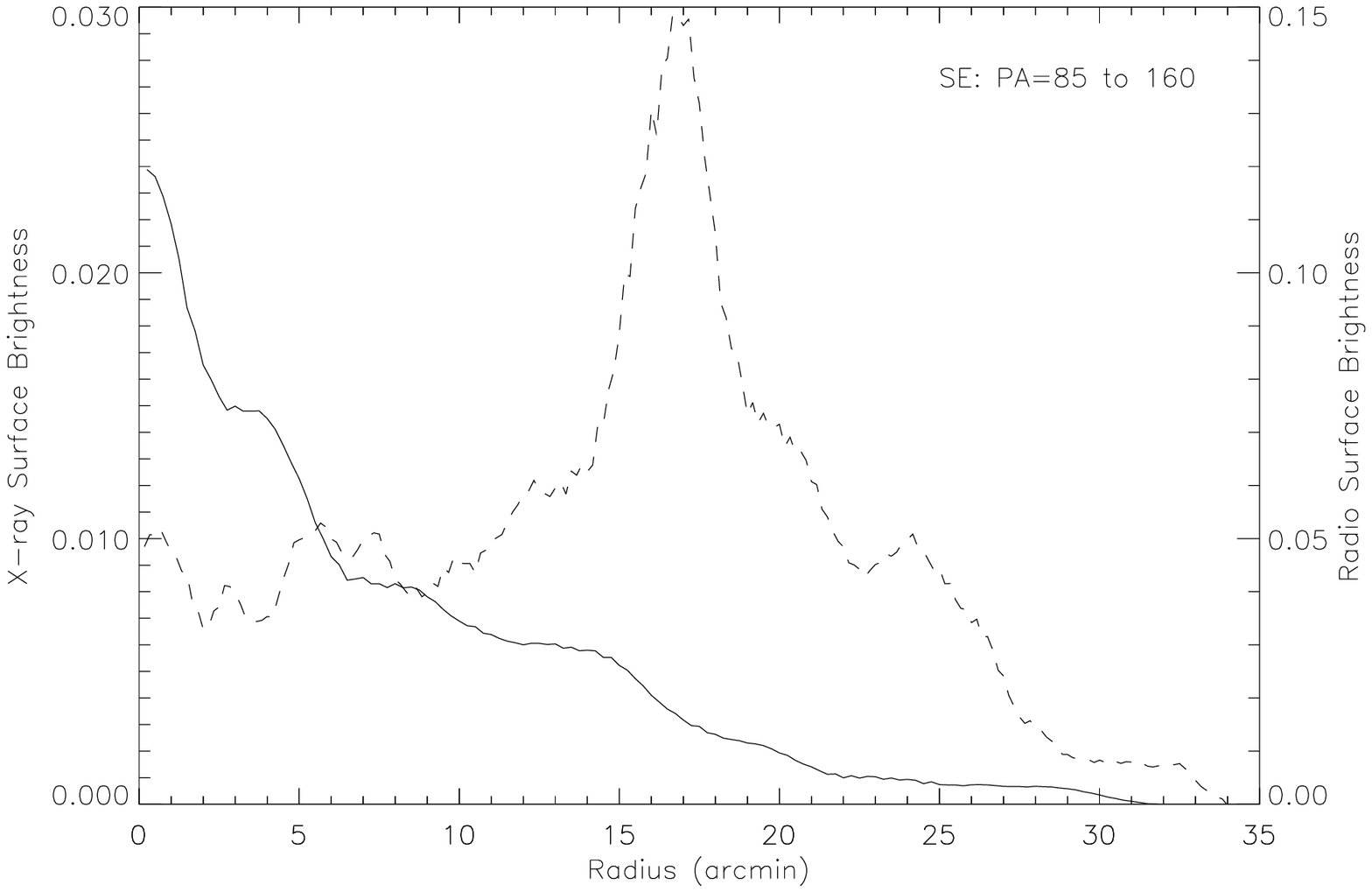,width=8truecm,height=7truecm}  
\psfig{figure=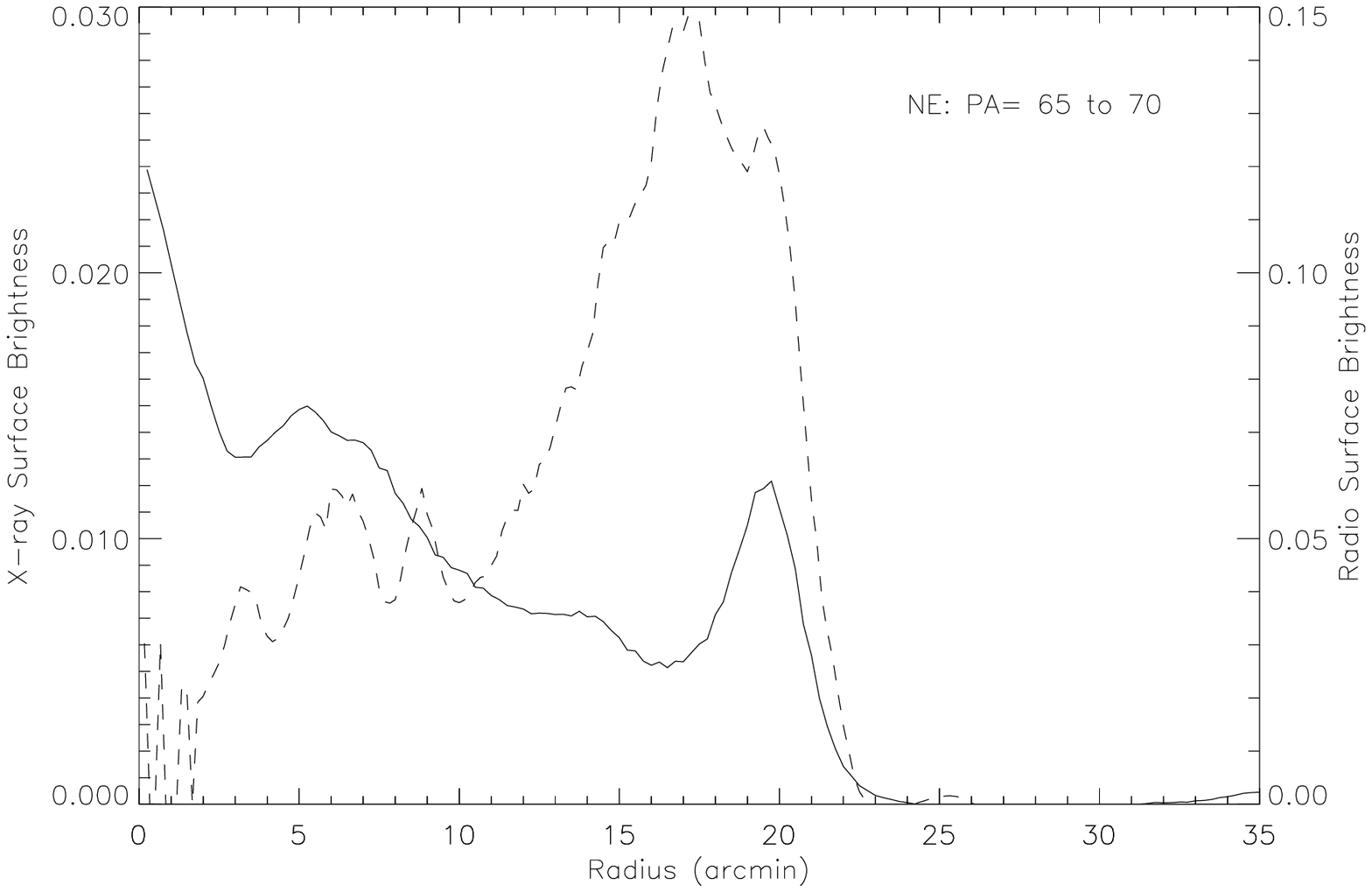,width=8truecm,height=7truecm}  
}}   
}  
\vskip 13truept  
{Fig. 9}

\psfig{figure=f10a.ps,height=7truecm,width=13truecm,angle=270}  
{Fig. 10a}  
  
\psfig{figure=f10b.ps,height=7truecm,width=13truecm,angle=270}  
{Fig. 10b}

\clearpage  
  
\psfig{figure=f10c.ps,height=8truecm,width=13truecm,angle=270}  
{Fig. 10c}  
\vskip 1.5truecm  
\psfig{figure=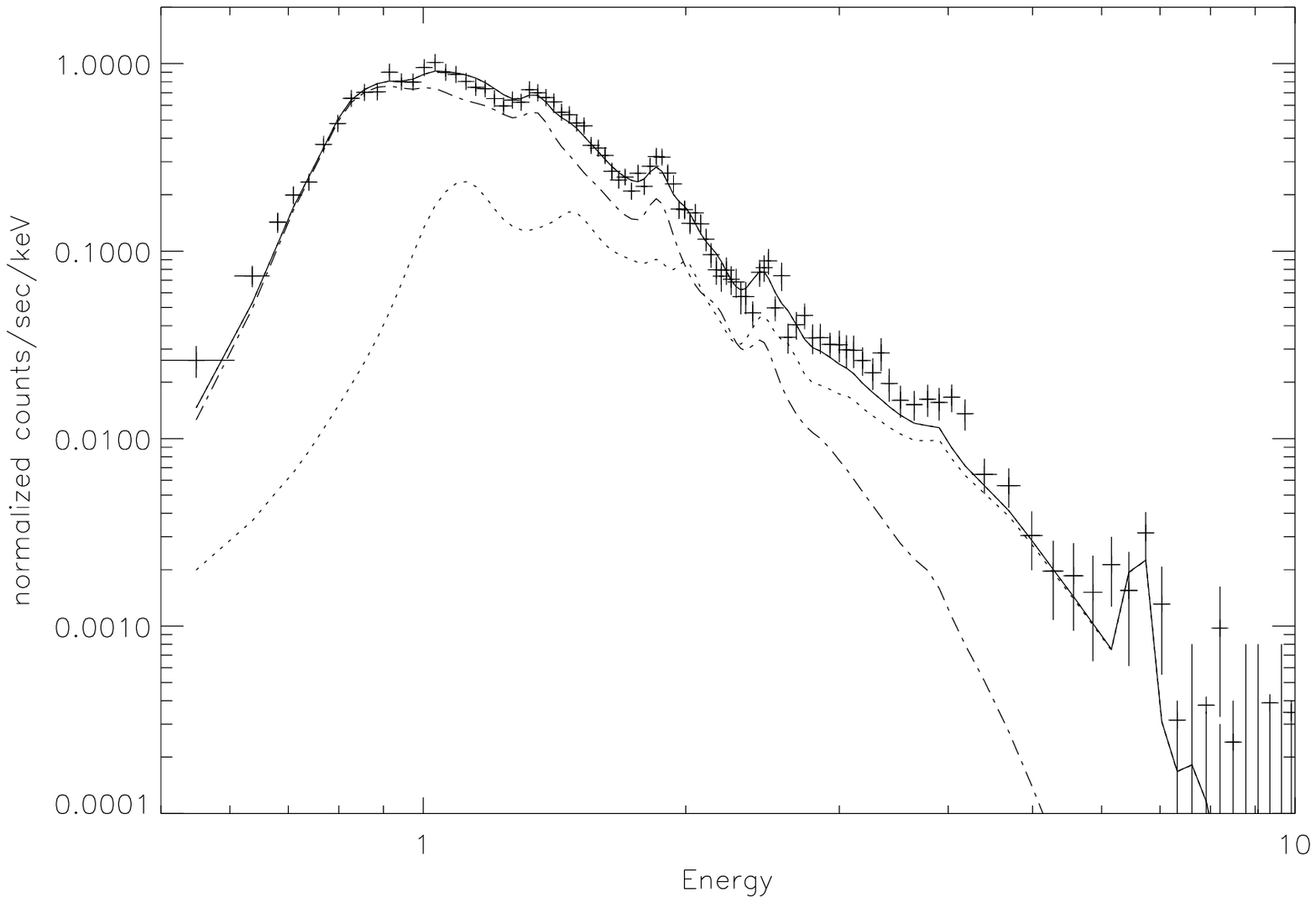,width=15truecm,height=8truecm}  
{Fig. 10d}

\clearpage  
{Fig. 11 is a jpeg file.}   

\end{document}